\documentclass[12pt,reqno]{amsart}
\usepackage{latexsym,amssymb}


\newcommand{\be}{\begin{equation}}
\newcommand{\ee}{\end{equation}}

\newcommand{\ba}{\begin{eqnarray}}
\newcommand{\ea}{\end{eqnarray}}

\newcommand{\no}{\nonumber}

\newcommand{\bma}{\begin{array}}
\newcommand{\ema}{\end{array}}

\newcommand{\ul}{\underline}
\newcommand{\ol}{\overline}
\newcommand{\lb}{\label}

\newcommand{\bbN}{{\Bbb{N}}}
\newcommand{\bbR}{{\Bbb{R}}}

\newcommand{\bbZ}{{\Bbb{Z}}}
\newcommand{\bbC}{{\Bbb{C}}}

\newcommand{\calC}{{\mathcal C}}

\newcommand{\calD}{{\mathcal D}}

\newcommand{\calM}{{\mathcal M}}
\newcommand{\calE}{{\mathcal E}}
\newcommand{\calL}{{\mathcal L}}
\newcommand{\calF}{{\mathcal F}}
\newcommand{\calK}{{\mathcal K}}

\newcommand{\bi}{\bibitem}

\newcommand{\f}{\frac}
\newcommand{\mb}{\mbox}
\newcommand{\uz}{{\underline{z}}}
\newcommand{\huxi}{{\hat{\underline{\Xi}}}}
\newcommand{\uxi}{{\underline{\Xi}}}
\newcommand{\lam}{\lambda}
\newcommand{\sig}{\sigma}
\newcommand{\eps}{\epsilon}
\newcommand{\gam}{\gamma}
\newcommand{\ome}{\omega}
\newcommand{\Ome}{\Omega}
\newcommand{\al}{\alpha}
\newcommand{\del}{\delta}
\newcommand{\pa}{\partial}

\newcommand{\tht}{\theta}

\renewcommand{\inf}{\infty}
\newcommand{\infpm}{{\infty_+, \infty_-}}

\newcommand{\ti}{\tilde  }

\newtheorem{prop}{Proposition}[section]
\newtheorem{thm}[prop]{Theorem}

\newtheorem{cor}[prop]{Corollary}
\newtheorem{lem}[prop]{Lemma}

\theoremstyle{definition}
\newtheorem{rem}[prop]{Remark} 

\theoremstyle{remark}
\newtheorem{exmp}[prop]{\bf Example}

\newcommand{\res}{\operatornamewithlimits{res}}

 \numberwithin{equation}{section}

\DeclareMathOperator{\Ker}{Ker}

\newcommand{\Div}{\operatorname{Div}}

\allowdisplaybreaks

\begin{document}

\title{AN ALTERNATIVE Approach to Algebro-Geometric
Solutions of the AKNS Hierarchy}

\author{F.~Gesztesy}
\address{Department of Mathematics, University of
Missouri, Columbia, MO 65211, USA.}
\email{mathfg@mizzou1.missouri.edu}
\author{R.~Ratnaseelan}
\address{Department of Mathematics, University of
Missouri, Columbia, MO 65211, USA.}
\email{ratnasr@towers.com}

\maketitle

\begin{abstract}
We develop an alternative systematic approach to the AKNS
hierarchy based on elementary algebraic methods. In particular, we
recursively construct Lax pairs for the entire AKNS hierarchy by
introducing a fundamental polynomial formalism and establish
 the basic algebro-geometric setting including associated
Burchnall-Chaundy curves, Baker-Akhiezer functions, trace formulas,
Dubrovin-type equations for analogs of Dirichlet and
Neumann divisors, and theta function representations for
algebro-geometric solutions.

Submitted to {\it Rev. Math. Phys.}
\end{abstract}

\section{Introduction} \label{s1}

The principal aim of this paper is an alternative
elementary algebraic approach to the entire AKNS hierarchy
in the spirit of previous treatments of the Korteweg-de Vries
(KdV), Boussinesq, and Toda hierarchies. More precisely, we
advocate a fundamental polynomial formalism to recursively
construct Lax pairs for the AKNS hierarchy, that is, pairs
$(D,E_{n+1})$ of matrix-valued differential expressions of order
one (i.e., $D$) and $n+1$ (i.e., $E_{n+1}$) with $D$ of the
Dirac-type. In addition, we establish the basic
algebro-geometric setup for special classes of solutions of the
AKNS hierarchy including solitons, rational solutions,
algebro-geometric quasi-periodic solutions, and limiting cases
thereof.
Our treatment includes a systematic approach to
Burchnall-Chaundy curves, Baker-Akhiezer functions, trace
formulas, Dubrovin-type equations describing the dynamics of
Dirichlet and Neumann divisors, and theta function
representations for algebro-geometric solutions.

Before we enter a description of the contents of each section,
it seems appropriate to comment on existing treatments of this
subject and to justify the addition of yet another detailed
account on this topic. The theory of commuting matrix-valued
differential expressions and, more generally, the
algebro-geometric approach to matrix hierarchies of soliton
equations has been developed in great generality by Dubrovin
and Krichever. Corresponding authoritative accounts can be found,
for instance, in \cite{bbei94}, Chs. 3, 4, \cite{dub77},
\cite{dub83}, \cite{dkn90}, \cite{its1}, \cite{its2}, \cite{kri1},
\cite{kri2}, \cite{pre85}, and the references therein.
In contrast to these references, our own approach relies
on two basic ingredients, an elementary polynomial approach
to Lax pairs (or zero-curvature pairs) of the AKNS hierarchy
and its explicit connection with a fundamental meromorphic
quantity $\phi$ (cf. (\ref{310}), (\ref{415})) which allows
for a unified algebro-geometric treatment of the entire AKNS
hierarchy.

In section 2 we describe a recursive approach to Lax pairs (and
zero-curvature pairs) of the AKNS hierarchy following Al'ber's
treatment of the KdV and nonlinear Schr\"odinger hierarchies
\cite{alb79}, \cite{alb81}, \cite{alb87} and establish its
connection with the Burchnall-Chaundy theory
\cite{bc1}, \cite{bc2}, \cite{bc3} and hence with hyperelliptic
curves. Combining the recursive formalism of Section 2 with a
polynomial approach to represent positive divisors of degree
$n$ of a hyperelliptic curve of genus $n$ originally developed
by Jacobi \cite{jac} and applied to the KdV case by Mumford
\cite{mum2}, Section III.a.1 and McKean \cite{mk85} (see also
\cite{ef}, \cite{pre96}), a detailed
analysis of the stationary AKNS hierarchy is provided in
Section 3. This includes, in particular, the theta function
representation of algebro-geometric solutions of the stationary
AKNS hierarchy. The corresponding time-dependent formalism is then
developed in detail in Section 4. Appendix A collects the relevant
material for hyperelliptic Riemann surfaces and their theta
functions. Appendix B contains a simple illustration of the
Riemann-Roch
theorem (cf. Theorem B.1).

We emphasize that our treatment comprises, in particular, the
important special case of the nonlinear Schr\"odinger (NS)
equation (cf. (\ref{372})), whose algebro-geometric solutions
have been studied, for instance, in \cite{bbei94}, Ch. 4,
\cite{dub83}, \cite{its1}, \cite{its2}, \cite{ma}, \cite{mer},
\cite{pre85}. Similarly, the case of the modified Korteweg-de
Vries (mKdV) equation (cf. (\ref{374}), whose algebro-geometric
solutions have been studied, for instance, in \cite{ges92},
\cite{gs}, are included as a special case of our formalism.
Moreover, the present elementary approach is not at all restricted
 to the AKNS hierarchy but applies quite generally to 1+1-dimensional
hierarchies of soliton equations. In fact,
the KdV case has been
treated in \cite{grt}, the case of the Toda and Kac-van Moerbeke
hierarchies in \cite{bght}, and the case of the Boussinesq
hierarchy in \cite{dgu}.

Finally, we mention that a combination of the AKNS formalism
developed in this paper and the Picard-type techniques introduced
in a recent explicit characterization of all elliptic solutions
of the KdV hierarchy \cite{geweAM} (see also \cite{geweMZ})
are expected to yield a similar characterization of all elliptic
solutions of the AKNS hierarchy, a topic that continues to attract
considerable interest (see, e.g., \cite{bbei94}, Ch. 7,
\cite{ceek95}, \cite{smi95}).


\section{The AKNS Hierarchy, Recursion Relations, and Hyperelliptic
Curves} \label{s2}

In this section we briefly review the construction of the AKNS
hierarchy
using a recursive approach advocated by Al'ber \cite{alb79},
\cite{alb81}, \cite{alb87}
(see also \cite{dickeyb}, Ch. 12, \cite{gd79},
\cite{gw93}, \cite{geweMN},  \cite{grt}) and outline its
 connection with the analog of the Burchnall-Chaundy
polynomial \cite{bc1}, \cite{bc2}, \cite{bc3},
and associated hyperelliptic curves.

Suppose $p,q \in C^{\infty}(\bbR)$ (or meromorphic on $\bbC$) and
introduce the Dirac-type matrix-valued differential expression
\ba
D = i \begin{pmatrix}
                  \f{d}{dx} & -q \\ p & -\f{d}{dx}
                   \end{pmatrix}, \; \; x\in \bbR\; \;(\mb{or}\; \;\bbC).
\label{21}
\ea
 In order to explicitly construct higher-order matrix-valued differential
 expressions $E_{n+1}$,    $n\in \bbN_0\; \;(=\bbN\cup\{0\})$ commuting
with $D$, which will be used to
define the stationary AKNS hierarchy
later, one can proceed as follows.

Pick $n\in \bbN_0$ and define
$ \left \{f_{\ell}(x)\right\}_{0 \leq \ell
\leq n},\; \; \left
\{g_{\ell}(x)\right\}_{0 \leq \ell \leq n+1},\; \;
\left\{h_{\ell}(x)\right\}_{0 \leq \ell \leq n }$ recursively by
\ba
&& f_0(x)=-iq(x),\; \;g_0(x)=1,\; \;h_0(x)=ip(x), \no \\
&& f_{\ell+1}(x)= \f{i}{2} f_{\ell,x}(x) - i q(x) g_{\ell+1}(x),
 \; \; 0 \leq \ell \leq n-1, \no \\
&& g_{\ell+1,x}(x) = p(x)f_{\ell}(x)
+ q(x)h_{\ell}(x) ,\; \; 0 \leq \ell \leq n, \label{22} \\
&& h_{\ell+1}(x)= -\f{i}{2} h_{\ell,x}(x) + i p(x) g_{\ell+1}(x),
 \; \; 0 \leq \ell \leq n-1. \no
\ea
Explicitly, one computes
\ba
&&f_0 = -iq,\no \\
&& f_1=\f12 q_x+c_1(-iq),\no \\
&&f_2 = \f{i}{4} q_{xx}-
      \f{i}{2}pq^2+ c_1( \f12 q_x)+c_2(-iq), \no \\
&&g_0 = 1,\no \\
&&g_1=c_1,\no \\
&& g_2=\f12 pq+c_2,\no \\
&&g_3 =  -\f{i}{4}(p_{_x}q - pq_x) + c_1(\f12 pq) + c_3 , \label{23}\\
&&h_0 =i p,\no \\
&&h_1=\f12 p_{_x}+ c_1(ip), \no \\
&&h_2=-\f{i}{4} p_{_{xx}}+\f{i}{2}p{^2}q+ c_1( \f12
p_{_x})+c_2(i p), \no \\ \no \\
&&\text{etc}., \no
\ea
where $\{c_\ell\}_{1 \leq \ell \leq n+1}$ are integration constants.
Given (\ref{22}), one defines the matrix-valued differential expression
$E_{n+1}$ by
\ba
E_{n+1}=i \sum_{\ell=0}^{n+1}\begin{pmatrix}
                  -g_{n+1-\ell} & f_{n-\ell} \\ -h_{n-\ell}
& g_{n+1-\ell}
                   \end{pmatrix}D^{\ell}, \; \; n\in \bbN_0,\;
\; f_{-1}=h_{-1}=0,
\label{24}
\ea
and verifies
\ba
[E_{n+1},D]= \begin{pmatrix}
                  0 & -2if_{n+1} \\ 2 i h_{n+1} & 0
                   \end{pmatrix}, \; \; n\in \bbN_0
\label{25}
\ea
(\ [ . , . ] the commutator symbol). The pair $(E_{n+1},D)$
represents the
 celebrated Lax pair for the AKNS hierarchy.
Varying $n \in \bbN_0$, the stationary AKNS hierarchy is then
defined by
the vanishing of the commutator of $E_{n+1}$ and $D$ in (\ref{25}),
that is, by
\ba
[E_{n+1},D]= 0,  \; \; n\in \bbN_0,
\label{26}
\ea
or equivalently, by
\ba
f_{n+1}=h_{n+1}=0, \; \; n\in \bbN_0.
\label{27}
\ea
Explicitly, one obtains for the first
few equations of the stationary AKNS hierarchy,
 \ba
&&\left\{ \bma{l}- p_{_x} + c_1(-2 i p) = 0, \\
           \\
        - q_x + c_1(2 i q) = 0, \ema \right. \no\\ \no \\
&&\left\{ \bma{l}\f{i}{2} p_{_{xx}}-i p{^2}q +
    c_1( - p_{_x})
    +c_2(-2 i p) = 0, \\ \label{28}\\
       -\f{i}{2} q_{xx}+i pq^2
    + c_1(-q_x)+c_2(2 i q)= 0, \ema \right. \\ \no \\
&&\left\{ \bma{l}\f14p_{_{xxx}}- \f32pp_xq +
    c_1(\f{i}2p_{_{xx}}- i p{^2}q)+
          c_2(-p_{_x})+c_3(-2ip) = 0, \\
           \\
    \f14q_{xxx}-\f32pqq_x +
    c_1(-\f{i}2q_{xx}+i pq^2)+c_2(-q_x)+c_3(2iq)= 0,
\ema \right. \no \\ \no \\
&&\text{etc}. \no
\ea

By definition, solutions $(p(x),q(x))$ of any of the
stationary AKNS
equations (\ref{28}) are called  {\bf algebro-geometric
stationary
finite-gap solutions} associated with the AKNS hierarchy. If
 $(p,q)$ satisfies
the $n^{\mb{\small th}}$ equation $(n \in \bbN_0)$ of
(\ref{28}) one also
calls $(p,q)$ a stationary $n$-gap solution.

Next, we introduce polynomials $F_n,\; \;G_{n+1}, \;
\;H_n$ with respect
to $z\in \bbC$,
\ba
&&F_n(z,x)= \sum_{\ell=0}^{n}f_{n-\ell}(x)z^{\ell},\; \;
f_0(x)=-iq(x), \no \\
&&G_{n+1}(z,x)= \sum_{\ell=0}^{n+1}g_{n+1-\ell}(x)z^\ell,\;
\; g_0(x)=1,
 \label{29}\\
&&H_n(z,x)= \sum_{\ell=0}^{n}h_{n-\ell}(x)z^\ell,\; \;
h_0(x)=ip(x). \no
\ea
and note that (\ref{26}) respectively, (\ref{27}) become
\ba
&&F_ {n,x}(z,x) =  -2 i z F_n(z,x) +2 q(x) G_{n+1}(z,x),
\label{210}\\
&&G_ {n+1,x}(z,x) =  p(x)F_n(z,x) + q(x)H_n(z,x),
\label{211} \\
&&H_ {n,x}(z,x) =  2 i z H_n(z,x) +2 p(x) G_{n+1}(z,x).
\label{212}
\ea
(\ref{210})--(\ref{212}) yield that
\ba
\left( G_{n+1}^2 - F_nH_n \right)_x =0
\label{213}
\ea
and hence
\ba
G_{n+1}(z,x)^2- F_n(z,x)H_n(z,x)=R_{2n+2}(z)  ,
\label{214}
\ea
where the integration constant $R_{2n+2}(z)$ is a monic
polynomial in $z$ of
degree $2n+2$. Thus one may write
\ba
R_{2n+2}(z)= \prod_{m=0}^ {2n+1} (z-E_m),\; \;
\{E_m\}_{0\leq m \leq 2n+1}
\subset \bbC.
\label{215}
\ea
Explicitly, one obtains for the first few polynomials in (\ref{29}),
 \ba
&&F_0=-iq, \no \\
&& F_1=-iqz+\f12 q_x + c_1(-iq), \no \\
&& F_2=-iqz^2+\f12 q_x z
    +\f{i}{4} q_{xx}- \f{i}{2}pq^2
  +c_1(-iqz+\f12q_x)+c_2(-iq), \no \\
&&G_1=z+c_1 , \no \\
&& G_2 = z^2 +\f12 pq+c_1 z+c_2, \no \\
&& G_3 = z^3 +\f12pqz-\f{i}{4}(p_xq
        -pq_x) +c_1(z^2+\f12pq) +c_2 z+c_3,  \label{216}\\
&&H_0=i p, \no \\
&&H_1=ipz+\f12 p_{_x}+ c_1(ip),\no \\
&& H_2=ipz^2+\f12 p_{_x} z -\f{i}{4} p_{_{xx}}+
    \f{i}{2}p{^2}q+c_1(ipz+\f12p_x)+c_2(ip), \no \\ \no \\
&&\text{etc}. \no
\ea

One can use (\ref{210})--(\ref{212}) and (\ref{214}) to
derive differential equations for $F_n$ and $H_n$ separately by
eliminating $G_{n+1}$. One obtains for $F_n$,
\ba
&& F_nF_{n,xx}-\f{q_x}{q}F_nF_{n,x}-\f12F_{n,x}^2+(2 z^2-2iz
\f{q_x}{q}-2pq)F_n^2 \no \\
&&=-2 q^2 R_{2n+2}(z)
\lb{216a}
\ea
and upon dividing (\ref{216a}) by $q^2$ and differentiating the
result with respect to $x$,
\ba
F_{n,xxx}-3\f{q_x}{q}F_{n,xx}+(4z^2-4iz\f{q_x}{q}-4pq-\f{q_{xx}}{q}
+3\f{q_x^2}{q^2})F_{n,x} \no\\
+(-4z^2\f{q_x}{q}+6iz\f{q_x^2}{q^2}-2iz\f{q_{xx}}{q}
+2pq_x-2p_xq)F_n=0.
\lb{216b}
\ea
Similarly one obtains for $H_n$,
\ba
&& H_nH_{n,xx}-\f{p_x}{p}H_nH_{n,x}-\f12H_{n,x}^2+(2 z^2+2iz
\f{p_x}{p}-2pq)H_n^2 \no \\
&&=-2 p^2 R_{2n+2}(z),
\lb{216c}
\ea
\ba
H_{n,xxx}-3\f{p_x}{p}H_{n,xx}+(4z^2+4iz\f{p_x}{p}-4pq-\f{p_{xx}}{p}
+3\f{p_x^2}{p^2})H_{n,x} \no\\
+(-4z^2\f{p_x}{p}-6iz\f{p_x^2}{p^2}+2iz\f{p_{xx}}{p}
+2p_xq-2pq_x)H_n=0.
\lb{216d}
\ea
(\ref{216a}) and (\ref{216c}) can be used to derive recursion
relations for $f_\ell$ and $h_\ell$ in the homogeneous case
where all
 $c_\ell=0,\; \;\ell\in\bbN$ (cf. Lemma 4.5). This has
interesting
applications to the high-energy expansion of the Green's
matrix of $D$ as briefly discussed in Remark \ref{r46}.

Next, we consider the kernel
(i.e., the formal null space in a purely algebraic
sense) of $(D-z), \; \; z \in \bbC$,
\ba
(D-z)\Psi=0,\; \; {\Psi(z,x)} =\begin{pmatrix}
                  {\psi}_1(z,x) \\ {\psi}_2(z,x)
                   \end{pmatrix}, \; \; z \in \bbC
\label{217}
\ea
and, taking into account (\ref{26}), that is, $[E_{n+1},D]=0$,
compute the
restriction of $E_{n+1}$ to $\mb{Ker}(D-z)$. Using
\ba
\psi_{1,x}=-iz\psi_1+q\psi_2,\; \psi_{2,x}=iz\psi_2+p\psi_1,
\; \;
\mb{etc.},
\label{218}
\ea
in order to eliminate higher-order derivatives of $\psi_j,\;
j=1,2$, one
obtains from (\ref{22}), (\ref{24}), (\ref{27}), (\ref{29}), and
(\ref{210})--(\ref{212}),
\ba
E_{n+1}\Big{\vert}_{\Ker(D-z)}= i\begin{pmatrix}
                  -G_{n+1}(z,x) & F_{n}(z,x) \\ -H_{n}(z,x)
& G_{n+1}(z,x)
                   \end{pmatrix} \Big{\vert}_{\Ker(D-z)} .
\label{219}
\ea
Still assuming $f_{n+1}=h_{n+1}=0$ as in (\ref{27}),
$[E_{n+1},D]=0$ in
 (\ref{26})
yields an algebraic relationship between $E_{n+1}$ and $D$
by a celebrated  result of Burchnall and Chaundy
\cite{bc1}, \cite{bc2}, \cite{bc3} (see also \cite{pre96},
\cite{wi85}).
The following theorem
details this relationship.

\begin{thm}\label{t21}
Assume $f_{n+1}=h_{n+1}=0$, that is, $[E_{n+1},D]=0$ for some
$n \in \bbN_0$.
Then the Burchnall-Chaundy polynomial $\calF_n(D,E_{n+1})$ of
the pair
 $(D,E_{n+1})$ explicitly reads (cf. (\ref{215}))
\ba
\calF_n(D,E_{n+1}) = E_{n+1}^2+R_{2n+2}(D)=0, \; \;
R_{2n+2}(z)
       = \prod_{m=0}^{2n+1} (z-E_m), \; \; z\in \bbC.
\label{220}
\ea
\end{thm}
\begin{proof}
$[E_{n+1},D]=0$, (\ref{214}), and (\ref{219}) imply
 \ba
&& E_{n+1}^2\Big{\vert}_{\Ker(D-z)} =
\left[E_{n+1}\Big{\vert}_{\Ker(D-z)}
        \right]^2\Big{\vert}_{\Ker(D-z)} \no \\
&&=-\begin{pmatrix}
                  G_{n+1}^2-F_nH_n & 0 \\{} 0 & G_{n+1}^2-F_nH_n
                   \end{pmatrix}  \Big{\vert}_{\Ker(D-z)}
   = -R_{2n+2}(z)\begin{pmatrix}
                  1 & 0 \\0 & 1
                   \end{pmatrix} \Big{\vert}_{\Ker(D-z)} \no \\
&&=-R_{2n+2}(D)\Big{\vert}_{\Ker(D-z)}.
\label{221}
\ea
Since $z \in \bbC$ is arbitrary one infers (\ref{220}).
\end{proof}

\begin{rem} \label{r22}
Equation (\ref{220}) naturally leads to the (possibly singular)
hyperelliptic
curve $\calK_n$,
\ba
\calK_n:\; \; \calF_n(z,y)=y^2-R_{2n+2}(z)=0, \; \;
R_{2n+2}(z)
       = \prod_{m=0}^{2n+1} (z-E_m), \; \;n\in \bbN_0 \label{222}
\ea
of (arithmetic) genus $n$.
\end{rem}

Next, introducing a deformation parameter
$t_n \in \bbR\; \; \mb{in}\; \; (p,q)\; \;(i.e.,\; (p(x),q(x))
\rightarrow
(p(x,t_n),q(x,t_n)))$,
the time-dependent
AKNS hierarchy (cf., e.g., \cite{new}, Chs. 3, 5 and
the references therein) is defined as the collection of
evolution equations
(varying $n \in \bbN_0$),
\ba
\f{d}{dt_n}D(t_n) - [E_{n+1}(t_n),D(t_n)]=0, \; \; (x,t_n)
\in \bbR^2, \; \;
             n\in \bbN_0,
\label{223}
\ea
or equivalently, by
\ba
\mb{AKNS}_n(p,q)&=&\left\{ \bma{l}p_{t_n}(x,t_n)-
2 h_{n+1}(x,t_n)=0, \\
\\
 q_{t_n}(x,t_n)- 2 f_{n+1}(x,t_n)=0, \ema \right.
 \; \;(x,t_n) \in \bbR^2, \; \; n \in \bbN_0,  \label{224}
\ea
that is, by
\ba
\mb{AKNS}_n(p,q)=\left\{ \bma{l}p_{t_n}+iH_{n,x}
+2zH_n-2ipG_{n+1}=0, \\
 \\
q_{t_n}-iF_{n,x}+2zF_n+2iqG_{n+1}=0, \ema \right.
\; \;(x,t_n) \in \bbR^2,\; \; n \in \bbN_0.
\label{225}
\ea
Explicitly, one obtains for the first few equations in
(\ref{224}) or
(\ref{225}),
\ba
&& \mb{AKNS}_0(p,q)=\left\{ \bma{l}p_{t_{0}}- p_{_x} +
 c_1(-2 i p) = 0,
       \no \\ \no  \\
    q_{t_{0}} - q_x + c_1(2 i q) = 0,     \ema \right.
\no \\ \no \\
&& \mb{AKNS}_1(p,q)=\left\{ \bma{l}
         p_{t_{1}}  +\f{i}{2} p_{_{xx}}-i p{^2}q +
    c_1\left( - p_{_x}\right)
    +c_2(-2 i p) = 0,  \\ \label{226}
        \\
      q_{t_{1}}-\f{i}{2} q_{xx}+i pq^2
                           + c_1\left(- q_x\right)+c_2(2 i q) =0,
   \ema \right.  \\ \no  \\
&&\mb{AKNS}_2(p,q)=\left\{ \bma{l}
      p_{t_{2}}+\f14p_{_{xxx}}- \f32pp_xq +
    c_1(\f{i}2p_{_{xx}}- i p{^2}q)+
          c_2(-p_{_x})+c_3(-2ip) = 0, \no \\
       \no    \\
       q_{t_{2}}+ \f14q_{xxx}-\f32pqq_x +
   c_1(-\f{i}2q_{xx}+i pq^2)+c_2(-q_x)+c_3(2iq) = 0,
    \ema \right. \no \\ \no \\
&&\text{etc}. \no
\ea

\begin{rem} \label{r23}
We chose to start by postulating the recursion relation
(\ref{22}) and
then developed the whole formalism based on (\ref{22}),
(\ref{24})--(\ref{26}). Alternatively one could have started from
\ba
(D-z)\Psi(P)=0,\; \; (E_{n+1}-iy(p))\Psi(P)=0,\; \; P=(z,y)
\in \calK_n
\backslash \{\infty_{\pm}\}
\label{227}
\ea
and obtained the recursion relation (\ref{22}) and the remaining
stationary
 results of this section as a consequence of (\ref{29}) and
(\ref{219}).
 Similarly, starting with
\ba
(D-z)\Psi(P,t_n)=0,\; \;
 \left(\f{\partial}{{\partial}_{t_{n}}}-E_{n+1}\right)\Psi(P,t_n)=0,
\; \; t_n \in \bbR,
\label{228}
\ea
one infers the time-dependent results (\ref{223})--(\ref{226}).
\end{rem}

\begin{rem} \label{r24}
Define
\ba
 U(z,x) = \begin{pmatrix}
                  -iz & q(x) \\ p(x) & iz
                   \end{pmatrix} , \; \;
V_{n+1}(z,x) = i \begin{pmatrix}
                  -G_{n+1}(z,x) & F_n(z,x) \\  -H_n(z,x) &
                    G_{n+1}(z,x)
                   \end{pmatrix} .
\label{229}
\ea
Then (\ref{219}) implies
\ba
-i\begin{pmatrix}
                  1 & 0 \\ 0 & -1
                    \end{pmatrix}[E_{n+1},D]\Big{\vert}_{\Ker(D-z)}
       = \left\{{-V_{n+1,x}(z)
 +[U(z),V_{n+1}(z)]}\right\}\Big{\vert}_{\Ker(D-z)}
\label{230}
\ea
and the stationary part of this section, being a consequence of
$[E_{n+1},D]=0$, can equivalently be based on the equation
\ba
-V_{n+1,x} +[U,V_{n+1}]=0.
\label{231}
\ea
In particular, the hyperelliptic curve $\calK_n$ in (\ref{222})
is then
obtained from the characteristic equation for $V_{n+1}(z,x)$,
\ba
&& \mb{det}[yI - V_{n+1}(z,x)] = y^2 - \mb{det}[V_{n+1}(z,x)]
\no \\
&& \quad = y^2 - G_{n+1}(z,x)^2 + F_n(z,x)H_n(z,x)=
y^2-R_{2n+2}(z)= 0.
\label{232}
\ea
Similarly, the time-dependent part (\ref{224})--(\ref{226}),
being
 based on the Lax equation  (\ref{223}), can equivalently be
developed
 from the zero-curvature equation
\ba
U_{t_{n}}-V_{n+1,x} +[U,V_{n+1}]=0.
\label{233}
\ea
\end{rem}
In fact, since the latter approach (\ref{233}) is almost
universally adopted
in the contemporary literature on the AKNS hierarchy, we
decided to recall its
proper origin in connection with the Lax pair $[E_{n+1},D]$
and based our
 treatment on matrix-valued differential expressions instead.


\section{The Stationary AKNS Formalism} \label{s3}

In this section we continue our discussion of the AKNS
hierarchy and
concentrate on the stationary case. Following \cite{grt},
where the
analogous treatment of the stationary KdV hierarchy can be found,
 we outline the connections between the polynomial approach
described in Section 2 and a fundamental meromorphic function
$\phi(P,x)$ defined on the hyperelliptic curve
$\calK_n$. Moreover, we discuss in detail the associated
stationary Baker-Akhiezer function $\Psi(P,x,x_0)$, the common
eigenfunction
of $D$ and $E_{n+1}$ (we recall that $[E_{n+1},D]=0$), and
associated
positive divisors of degree $n$ on $\calK_n$ (which should
be considered
as the analogs of Dirichlet and Neumann divisors in the KdV
context).

We recall the hyperelliptic curve (\ref{222}),
\ba
\calK_n:\; \; \calF_n(z,y)=y^2-R_{2n+2}(z)=0,\; \;
\quad R_{2n+2}(z)
       = \prod_{m=0}^{2n+1} (z-E_m), \label{31}
\ea
where $n\in \bbN_0$ will be fixed throughout this section and
denote
its compactification (adding the points $\infty_{\pm}$) by the
same symbol.
Thus $\calK_n$ becomes a (possibly singular) two-sheeted
hyperelliptic
Riemann surface of arithmetic genus $n$ in a standard manner.
We shall
 introduce a bit more notation in this context (see Appendix A
for more
details). Points $P$ on $\calK_n$
are represented as pairs $P=(z,y)$ satisfying (\ref{31})
together with
$\infty_{\pm}=(\infty,\pm\infty)$, the points at infinity. The
complex structure
on $\calK_n$ is defined in the usual way by introducing local
coordinates
$\zeta_{P_0}: P \rightarrow (z-z_0)$ near points $P_0 \in
\calK_n$ which
are neither branch nor singular points of $\calK_n,\;
\zeta_{\infty_{\pm}}:
P \rightarrow \f1z \text{ near }\; \inf_{\pm}$, and similarly at
 branch and/or singular points of $\calK_n$.
 The holomorphic sheet exchange map (involution) $*$ is defined by
\ba
* : \left\{ \begin{array}{l}
{\calK}_n \to {\calK}_n\\
P=(z,y)
\mapsto P^*=(z,-y).
 \end{array} \right.
\label{32}
\ea
A detailed description of $\calK_n$ and its
complex structure
in the two most frequently discussed cases in applications
where either
$E_m \in \bbR,\; \;0 \leq m \leq 2n+1$ or
$\{E_m\}_{0 \leq m \leq 2n+1}=
\{E_{2m'},\ol{E_{2m'}}\}_{0 \leq m' \leq n}$ is provided
 at the end of Appendix A.

Finally, positive divisors on $\calK_n$ of degree are denoted by
\ba
{\calD}_{\ul{Q}} : \left\{ \begin{array}{l}
{\calK}_n \to \bbN_0\\
P \mapsto  {\calD}_{\ul{Q}}(P)\left\{ \begin{array}{ll}
m & \mb{if }
P \mb{ occurs } m \mb{ times in } \{Q_1,\ldots, Q_n\},\\ \\
0 & \mb{if } P \notin \{Q_1,\ldots, Q_n\},
 \end{array} \right.
\end{array} \right.\; \; {\ul{Q}}=(Q_1,\ldots, Q_n).\no
\ea
\ba
\label{33}
\ea
Given these preliminaries, let $\Psi(P,x,x_0)$ denote the
common normalized
eigenfunction of $D$ and $E_{n+1}$, whose existence follows from the
commutativity of $D$ and $E_{n+1}$ (cf., eg., \cite{bc1}, \cite{bc2}
in the case of scalar differential expressions), that is, due to
\ba
[E_{n+1},D]=0,
\label{34}
\ea
for a given $n\in \bbN_0$, or equivalently, due to the requirement,
\ba
f_{n+1}=h_{n+1}=0.
\label{35}
\ea
Explicitly, this yields
\ba
D\Psi(P,x,x_0)=z\Psi(P,x,x_0),\; \;
E_{n+1}\Psi(P,x,x_0)=iy(P)\Psi(P,x,x_0), \no \\
\label{36} \\
\Psi(P,x,x_0)=\begin{pmatrix}
                  {\psi}_1(P,x,x_0) \\ {\psi}_2(P,x,x_0)
                   \end{pmatrix},\; \;
 P=(z,y)\in\calK_n \backslash \{\infty_{\pm}\},
\; \; x\in \bbR \no
\ea
for some fixed $x_0\in \bbR$ with the assumed normalization,
\ba
{\psi}_1(P,x_0,x_0)=1,\; \; P \in \calK_n \backslash
\{\infty_{\pm}\}.
\label{37}
\ea
$\Psi(P,x,x_0)$ is called the Baker-Akhiezer (BA) function.
Closely related
to $\Psi(P,x,x_0)$ is the following meromorphic function
$\phi(P,x)$ on
$\calK_n$, defined by
\ba
\phi(P,x)=\f{{\psi}_2(P,x,x_0)}{{\psi}_1(P,x,x_0)}, \; \;
P \in \calK_n,\; \; x\in \bbR.
\label{38}
\ea
Since $\phi(P,x)$ will be the fundamental object for the stationary
AKNS hierarchy, we next seek its connection with the recursion
formalism
of Section 2. Recalling (\ref{219}), one infers
\ba
E_{n+1}\Psi= i\begin{pmatrix}
                 F_n\psi_2-G_{n+1}\psi_1 \\ G_{n+1}\psi_2-H_n\psi_1
                   \end{pmatrix} =
iy\begin{pmatrix}
                 \psi_1 \\ \psi_2
                   \end{pmatrix}
\label{39}
\ea
and hence by (\ref{38}),
\ba
\phi(P,x)=\f{y(P)+G_{n+1}(z,x)}{F_n(z,x)}=\f{-H_n(z,x)}
{y(P)-G_{n+1}(z,x)},
\; \;P=(z,y) \in \calK_n.
\label{310}
\ea
By (\ref{29}) we may write,
\ba
F_n(z,x)= -iq(x)\prod_{j=1}^{n} (z-\mu_j(x)),
\label{311}
\ea
\ba
H_n(z,x)= ip(x)\prod_{j=1}^{n} (z-\nu_j(x)).
\label{312}
\ea
Defining
\begin{align}
\hat{\mu}_j (x)
&=({\mu}_j (x),G_{n+1}({\mu_j }(x),x))\in \calK_n,
 \, \, \, 1 \leq j \leq n, \; \; x\in \bbR,  \label{313}\\
\hat{\nu}_j (x)
&=({\nu_j }(x),-G_{n+1}({\nu_j }(x),x))\in \calK_n,
 \, \, \, 1 \leq j \leq n, \; \; x\in \bbR,   \label{314}
\end{align}
one infers from (\ref{310}) that the divisor $(\phi(P,x))$ of
$\phi(P,x)$
is given by
\ba
&&\left(\phi(P,x)\right)={\calD}_{\hat{\underline \nu}(x)}(P) -
{\calD}_{\hat{\underline \mu}(x)}(P)+{\calD}_{\infty_{+}}(P) -
{\calD}_{\infty_{-}}(P), \label{315} \\
&&\hat{\underline \nu} (x) = ( \hat \nu_1 (x), \ldots,
\hat \nu_n (x)),\;
\hat{\underline \mu}(x) =
(\hat \mu_1 (x), \ldots, \hat \mu_n (x)). \no
\ea
Here we used our convention (\ref{33}) and additive notation
for divisors.
 Equivalently, $\infty_{+},\hat \nu_1 (x), \ldots,
\hat \nu_n (x)$, are the $n+1$ zeros of $\phi(P,x)$ and
$\infty_{-},\hat \mu_1 (x), \ldots,
\hat \mu_n (x)$, its $n+1$ poles. Clearly $\mu_j (x)$ and
$\nu_j (x)$ play the analogous role of Dirichlet and Neumann
eigenvalues when comparing to the KdV case. In particular,
${\calD}_{\hat{\underline \mu}(x)}$ and
 ${\calD}_{\hat{\underline \nu}(x)}$ represent the corresponding
 analogs of Dirichlet and Neumann divisors.

Next we summarize a variety of properties of $\phi(P,x)$ and
 $\Psi(P,x,x_0)$.
\begin{lem} \label{l31}
Assume (\ref{34})--(\ref{38}), $P=(z,y) \in \calK_n \backslash
\{\infty_{\pm}\},$ and let $(z,x,x_0) \in \bbC \times \bbR^2$.
Then
\ba
& (i).&\; \; \Psi(P,x,x_0) \text{satisfies the first-order system
 (cf. (\ref{229})}) \no \\
&&\quad \quad \quad{\Psi}_{x}(P,x,x_0)= U(z,x){\Psi}(P,x,x_0),
  \label{318} \vspace{2mm}\\
&&\quad \, iy(P){\Psi}(P,x,x_0)= V_{n+1}(z,x){\Psi}(P,x,x_0).
  \label{318a} \vspace{2mm}\\
&(ii).& \; \; {\phi}(P,x) \; \;
         \mb{satisfies the Riccati-type equation}  \no \vspace{2mm}\\
&& \quad \quad \quad \phi_x(P,x)+q(x)\phi(P,x)^2-2iz\phi(P,x)=p(x).
    \label{319} \vspace{2mm}\\
&(iii).& \; \; \phi(P,x)\phi(P^*,x)=
  \f{H_{n} (z, x)}{F_{n} (z, x)}. \label{320} \\
&(iv).& \; \; \phi(P,x)+\phi(P^*,x)=
    \f{2G_{n+1} (z, x)}{F_{n} (z, x)}. \label{321} \\
&(v).& \; \; \phi(P,x)-\phi(P^*,x)= \f{2 y(P)}
    {F_{n} (z, x)}. \label{322} \\
&(vi).& \; \; {\psi}_1(P,x,x_0)= \exp\left\{\int_{x_0}^x \;dx'
      [-iz+q(x')\phi(P,x')] \right\}  \label{323}\\
& &\quad \quad \quad \quad \quad   = \left[\f{F_n (z, x)}{F_n
    (z, x_0)}\right]^{\f12}
 \exp\left\{ y(P)
\int_{x_0}^x \;dx' q(x')F_n (z, x')^{-1}\right\}.  \label{324}\\
&(vii).& \; \; \psi_1(P,x,x_0)\psi_1(P^*,x,x_0)=
  \f{F_{n} (z, x)}{F_{n} (z, x_0)}. \label{325} \\
&(viii).& \; \; \psi_2(P,x,x_0)\psi_2(P^*,x,x_0)=
  \f{H_{n} (z, x)}{F_{n} (z, x_0)}. \label{326}\\
&(ix).& \; \; \psi_1(P,x,x_0)\psi_2(P^*,x,x_0)+
    \psi_1(P^*,x,x_0)\psi_2(P,x,x_0) \no \\
&&\quad\quad\quad=\f{2 G_{n+1}(z,x)}{F_{n}(z,x_0)}.\label{326a}
\ea
\end{lem}

\begin{proof}
$(i)$ is an immediate consequence of (\ref{229}), (\ref{36}),
and (\ref{310}).
$(ii)$ follows from $(i)$, (\ref{218}) and (\ref{38}).
$(iii)$--$(v)$
are clear from (\ref{310}). (\ref{323}) follows from
(\ref{218}) and
(\ref{38}). (\ref{324}) is a consequence of $(iv),\; (v)$,
(\ref{210}),
(\ref{323}),
and
\ba
\phi(P)=\f12[\phi(P)+\phi(P^*)]+\f12[\phi(P)-\phi(P^*)] \no \\
=\f{G_{n+1}}{F_n}+\f{y}{F_n}=\f{1}{q}\left(\f{F_{n,x}}
{F_n}+iz \right)
 + \f{y}{F_n}.
 \label{327}
\ea
$(vii)$ is clear from (\ref{324}) and $(viii)$ is
a consequence of (\ref{38}), $(iii)$, and $(vii)$. Finally,
$(ix)$ is a consequence of  (\ref{38}), (\ref{321}), and
(\ref{325}).
\end{proof}

In order to motivate our introduction of the basic quantity
$\phi(P,x)$
we started with the common eigenfunction $\psi(P,x,x_0)$ of
$D$ and
$E_{n+1}$. However, given (\ref{214}) we could have defined
$\phi(P,x)$
as in (\ref{310}) and then verified that
$\Psi= \begin{pmatrix}
        \psi_1 \\ \psi_2
     \end{pmatrix}$
defined by (\ref{38}) and (\ref{323}) satisfies (\ref{36}) and
(\ref{37}).

Concerning the dynamics of the zeros $\mu_j(x)$ and $\nu_j(x)$ of
$F_n(z,x)$ and $H_n(z,x)$ one obtains the following Dubrovin-type
equations.

\begin{lem} \label{l32}
Assume (\ref{34})--(\ref{38}), (\ref{311}), (\ref{312}) and let
$x \in \bbR$. Then
\ba
&& (i).\quad \mu_{j,x}(x)  = \f{-2iy({\hat{\mu}_j }(x))}
{\prod_{\stackrel{\scriptstyle{k=1}}{\scriptstyle{k \not=j}}}^{n}
(\mu_{j}(x) - \mu_{k}(x))},
\quad 1 \leq j \leq n. \label{328} \\
&& (ii).\quad \nu_{j,x}(x) = \f{-2iy ({\hat\nu_j }(x))}
{\prod_{\stackrel{\scriptstyle{k=1}}{\scriptstyle{k \not=j}}}^{n}
(\nu_{j}(x) - \nu_{k}(x))},
\quad 1 \leq j \leq n.
\label{329}
\ea
\end{lem}
\begin{proof}
Combine (\ref{210}), (\ref{311}), and (\ref{313}) and (\ref{212}),
 (\ref{312}), and (\ref{314}) in order to arrive at (\ref{328}) and
(\ref{329}), respectively.
\end{proof}

Combining the polynomial approach of Section 2 with (\ref{311})
and
(\ref{312}) readily yields trace formulas for the AKNS
invariants. We
 indicate the first few of these below.

\begin{lem} \label{l33}
Assume (\ref{34})--(\ref{38}) and let $x\in \bbR$. Then
 \begin{align}
(i).\ \ & i\f{p_x(x)}{p(x)} - 2c_1=  2\sum_{j_1=1}^{ n}
\nu_{j_1}(x), \no\\
& \f14\f{p_{xx}(x)}{p(x)} -\f12 p(x)q(x)
+c_1 \left(\f{i}{2}\f{p_{x}(x)}{p(x)}
 \right)
-c_2=-\sum_{\substack{j_1,j_2=1\\ j_1 <j_2} }^{  n} \nu_{j_1}(x) \,
\nu_{j_2}(x), \label{330}\\
 &\text{etc.}  \no \\
(ii).\ \  & i\f{q_x(x)}{q(x)} + 2c_1= - 2\sum_{j_1=1}^{ n}
\mu_{j_1}(x), \no\\
 & \f14\f{q_{xx}(x)}{q(x)} -\f12 p(x)q(x)
+c_1 \left(-\f{i}{2}\f{q_{x}(x)}{q(x)}
 \right)
-c_2=-\sum_{\substack{j_1,j_2=1\\ j_1 <j_2} }^{  n} \mu_{j_1}(x) \,
\mu_{j_2}(x), \label{331} \\
 &\text{etc.}  \no
\end{align}
Here
 \begin{align}
&c_1  = -\f{1}{2} \sum_{m_1=0}^{2\, n+1} E_{m_1}, \no\\
&c_2 = \f{1}{2} \sum_{\substack{m_1,m_2=0\\
m_1 <m_2} }^{2\, n+1} E_{m_1} E_{m_2}-
 \f{1}{8} \Big(\sum_{m_1=0}^{2\, n+1} E_{m_1}\Big)^2,
\label{332} \\
&\text{etc.} \no
 \end{align}
\end{lem}

\begin{proof}
(\ref{330}) and (\ref{331}) follow by comparison of powers of
$z$ substituting
(\ref{311}) into (\ref{29}) taking into account (\ref{23}).
(\ref{332}) follows
in exactly the same way from (\ref{23}), (\ref{29}), and (\ref{312}).
\end{proof}

Finally, we shall provide an explicit representation of
$\Psi,\; \phi,\; p,$ and $q$ in terms of the Riemann
theta function associated with $\calK_n$. We freely employ
the notation established in Appendix A. In order to avoid the
trivial case $n=0$ (considered in Example \ref{e37}) we assume
$n\in\bbN$
 for the remainder of this argument.

Assuming $\calK_n$
to be nonsingular, that is,
\ba
E_m \neq E_{m'} \; \; \text{for} \; \;
0 \leq m, m'\leq 2\; n+1,
\lb{333}
\ea
we choose, without loss of generality, the base point
$P_0=(E_0,0)$ and denote by
$\ul A_{P_0}(.),\; \ul{\al}_{P_0}(.)$ the Abel maps
as defined in (\ref{a30})--(\ref{a32}),
and define $\huxi_{P_0}$, the vector of
Riemann constants, by
\ba
&&\uxi_{P_0}=\huxi_{P_0}\; \; (\text{mod}\; \;L_n), \no
\\ \lb{334} \\
&&\huxi_{P_0}
=(\hat\Xi_{P_0, 1}, \ldots, \hat\Xi_{P_0, n} ), \quad
\hat\Xi_{P_0, j}
= \left[ \f{1+\tau_{j,j}}2\right.
-\sum_{\stackrel{\scriptstyle{k=1}}
{\scriptstyle{k \not=j}}}^{n}
\left.\int_{a_k} \hat A_{P_0,j} \ome_k\right].
\no
\ea
Next, consider the normal differential of the third
kind
$\ome^{(3)}_{\inf_+, \inf_-}$, which has simple poles
at $\inf_+$ and
$\infty_-$, corresponding residues $+1$ and $-1$,
vanishing $a$-periods,
and is holomorphic otherwise on $\calK_n$.  Hence we have
(cf.\ (\ref{a36}), (\ref{a37}))
\begin{align}
& \ome^{(3)}_{\inf_+, \inf_-} =
\f{\prod_{j=1}^n (\ti\pi -\lam_j)
d\ti\pi}{y},
\quad \ome_{\inf_-, \inf_+}^{(3)}
=-\ome^{(3)}_{\inf_+, \inf_-},
\lb{335}\\
& \int_{a_j} \ome_{\inf_+, \inf_-}^{(3)} = 0,
\quad 1\leq j\leq n,
\lb{336}
\end{align}
\begin{align}
&\ul U^{(3)} =(U_1^{(3)},\ldots ,U_n^{(3)}),\no \\
&U_j^{(3)} = \f{1}{2\pi i}
\int_{b_j} \ome_{\inf_+, \inf_-}^{(3)}
=\hat A_{\inf_-, j} (\inf_+)
   = 2 \hat A_{P_0, j}
(\inf_+), \quad 1 \leq j \leq n, \lb{337}
\end{align}
\begin{align}
\int_{P_0}^P \ome_{\inf_+, \inf_-}^{(3)}
\underset{\zeta \to 0}{=}\pm [\ln(\zeta)-\ln(\ome_0)
+O(\zeta)], \; \; P=(\zeta^{-1},y)\; \;
\mb{near}\; \; \inf_{\pm},
\lb{338}
\end{align}
where the numbers $\{ \lam_j \}_{1\leq j \leq n}$
are determined by the
normalization (\ref{336}). The Abelian differential of the
second kind $\ome_{\inf_{\pm},0}^{(2)}$ (cf.\ (\ref{a34}),
(\ref{a35})) are chosen such that
\begin{equation}
\ome_{\infty_\pm,0}^{(2)}
\underset{\zeta \to 0}{=}[\zeta^{-2} +O(1)]
\, d\zeta \text{ near }
\infty_\pm,
\lb{339}
\end{equation}
\begin{equation}
 \int_{a_j} \ome_{\inf_{\pm},0}^{(2)} = 0,
\quad 1\leq j\leq n,
\lb{340}
\end{equation}
\begin{align}
&\ul U_0^{(2)} =(U_{0,1}^{(2)},\ldots, U_{0,n}^{(2)}),\; \;
U_{0,j}^{(2)} = \f{1}{2\pi i}
\int_{b_j} \Ome_{0}^{(2)},\quad
\Ome_{0}^{(2)}=\ome_{\inf_{+,0}}^{(2)}-\ome_{\inf_{-,0}}^{(2)},
\lb{341}
\end{align}
\begin{align}
\int_{P_0}^P \Ome_{0}^{(2)}
\underset{\zeta \to 0}{=}\mp [\zeta^{-1}+e_{0,0}
+e_{0,1}\zeta + O(\zeta^2)], \; \; P=(\zeta^{-1},y)\; \;
\mb{near}\; \; \inf_{\pm}.
\lb{342}
\end{align}

Next, we formulate the following auxiliary result.

\begin{lem} \lb{l34}
Let $\psi(.,x),\; x\in\bbR$ be
meromorphic on ${\calK}_n\backslash
\{\infty_+, \infty_-\}$ with essential singularities
at $\infty_\pm$ such
that $\ti \psi(.,x)$ defined by
\begin{equation}
\ti \psi (.,x) =\psi (.,x)
\exp \Big[ -i(x-x_0) \int_{P_0}^P
\Ome_{0}^{(2)}\Big]
\lb{342a}
\end{equation}
is multi-valued meromorphic on ${\calK}_n$ and its
divisor satisfies
\begin{equation}
(\ti \psi (.,x))\geq -{\calD}_{\hat{\ul \mu}(x)}.
\lb{342b}
\end{equation}
Define a divisor ${\calD}_0 (x)$ by
\begin{equation}
(\ti \psi (.,x))={\calD}_0 (x)
-{\calD}_{\hat{\ul \mu} (x)}.
\lb{342c}
\end{equation}
Then
\begin{equation}
{\calD}_0 (x) \in\sig^n {\calK}_n, \; {\calD}_0 (x) > 0,
\; \deg ({\calD}_0 (x))
=n.
\lb{342d}
\end{equation}
Moreover, if ${\calD}_0 (x)$ is nonspecial for all
$x\in\bbR$, that is, if
\begin{equation}
i ({\calD}_0 (x) ) =0, \; \;  x \in \bbR,
\lb{342e}
\end{equation}
then $\psi (.,x)$ is unique up to a constant
multiple (which may depend
on $x$).
\end{lem}
\begin{proof}
By the Riemann-Roch theorem (see (\ref{a42}))
there exists at least one
such function $\psi(.,x)$.  Suppose $\psi_j (.,x)$,
$j=1,2$ are two such
functions satisfying (\ref{342c}) with corresponding
divisors
${\calD}_{0,j} (x)$, $j=1,2$. Then
\begin{equation}
(\psi_1 (.,x) / \psi_2 (.,x)) = {\calD}_{0,1} (x)
-{\calD}_{0,2} (x).
\lb{342f}
\end{equation}
Since $i({\calD}_{0,2}(x))=0$, $\deg ({\calD}_{0,2}(x))
=n$ by hypothesis, the multi-valued version of
(\ref{a42}) yields $r(-{\calD}_{0,2}(x))=1,\; \; x\in\bbR$
and hence $\psi_1
(.,x)/ \psi_2 (.,x)$ is a constant on ${\calK}_n$.
(For simplicity of notation, we did not prescribe the path
of integration in (\ref{342a}). This in turn forced us
to use the multi-valued form of the Riemann-Roch theorem,
see, e.g., \cite{fk}, Sect. III.9.)
\end{proof}

Assuming $\calD_{\ul Q}$ to be nonspecial, that is ,
$i(\calD_{\ul Q})=0,\; \; \ul Q=(Q_1,\ldots,Q_n)$, a
special case of Riemann's vanishing theorem yields that
\begin{equation}
\tht(\uxi_{P_0}-\ul A_{P_0}(P)+
{\ul \al}_{P_0}(\calD_{\ul Q}))=0\; \;
\text{if and only if}\; \;
P\in \{Q_1,\ldots,Q_n\}.
\lb{343}
\end{equation}
Hence the divisor (\ref{315}) of $\phi(P,x)$ suggests
considering expressions of the type
\begin{equation}
C(x)\f{\tht(\uxi_{P_0}-\ul A_{P_0}(P)+
{\ul \al}_{P_0}(\calD_{\hat {\ul \nu}(x)}))}
{\tht(\uxi_{P_0}-\ul A_{P_0}(P)+
{\ul \al}_{P_0}(\calD_{\hat {\ul \mu}(x)}))}
\exp\left[\int_{P_0}^P\ome_{\inf_+,\inf_-}^{(3)}\right],
\lb{344}
\end{equation}
where $C(x)$ is independent of $P\in\calK_n$.
In fact, abbreviating
\ba
&& \uz(P,\ul Q)=\ul A_{P_0}(P)-{\ul \al}_{P_0}(\calD_{\ul Q})-
\uxi_{P_0}, \lb{345} \\
&&\uz_{\pm}(\ul Q)=\uz(\inf_{\pm},\ul Q),\; \;
\ul Q=(Q_1,\ldots,Q_n),
\lb{346}
\ea
one obtains the following theta function representation
 for $\phi,\; \Psi$ and $(p,q)$ (the analog of the celebrated
Its-Matveev formula \cite{itma} in the KdV context).

\begin{thm}\lb{t35}
Let $P\in\calK_n\backslash\{\infpm\},\; (x,x_0)\in\bbR^2$,
and assume $\calK_n$ to be nonsingular, that is,
$E_m \neq E_{m'}\text{ for } m \neq m',\;
0 \leq m, m'\leq 2n+1.$ Moreover, suppose
$\calD_{\hat {\ul \mu}(x)}$, or equivalently,
$\calD_{\hat {\ul \nu}(x)}$
to be nonspecial, that is, $i(\calD_{\hat {\ul \mu}(x)})=
i(\calD_{\hat {\ul \nu}(x)})=0.$ Then
\ba
&&\phi(P,x)=\f{2i}{q(x_0)\ome_0}
\f{\tht(\uz_-(\hat {\ul \mu}(x_0)))}
{\tht(\uz_+(\hat {\ul \mu}(x_0)))}
\f{\tht(\uz_+(\hat {\ul \mu}(x)))}{\tht(\uz_-(\hat {\ul \nu}(x)))}
\f{\tht(\uz(P,\hat {\ul \nu}(x)))}{\tht(\uz(P,\hat {\ul \mu}(x)))}
\times \no \\
&&\quad\quad\quad
\times\exp\left[\int_{P_0}^P\ome_{\inf_+, \inf_-}^{(3)}
-2i(x-x_0)e_0\right],
\lb{347}\vspace{2mm} \\
&&\psi_1(P,x,x_0)=
\f{\tht(\uz_+(\hat {\ul \mu}(x_0)))}
{\tht(\uz(P,\hat {\ul \mu}(x_0)))}
\f{\tht(\uz(P,\hat {\ul \mu}(x)))}{\tht(\uz_+(\hat {\ul \mu}(x)))}
\exp\left[i(x-x_0)\left(e_0+\int_{P_0}^P\Ome_0^{(2)}\right)\right],
\quad\quad  \lb{348}\vspace{2mm} \\
&&\psi_2(P,x,x_0)=\f{2i}{q(x_0)\ome_0}
\f{\tht(\uz_-(\hat {\ul \mu}(x_0)))}{\tht(\uz(P,
\hat {\ul \mu}(x_0)))}
\f{\tht(\uz(P,\hat {\ul \nu}(x)))}{\tht(\uz_-(\hat {\ul \nu}(x)))}
\times\no \\
&&\quad\quad\quad\times\exp\left[\int_{P_0}^P\ome_{\inf_+,
\inf_-}^{(3)}+
i(x-x_0)\left(-e_0+\int_{P_0}^P\Ome_0^{(2)}\right)\right].
\lb{349}
\ea
Moreover, one derives
\ba
&&p(x)=p(x_0)
\f{\tht(\uz_-(\hat {\ul \nu}(x_0)))}{\tht(\uz_+(\hat {\ul
\nu}(x_0)))}
\f{\tht(\uz_+(\hat {\ul \nu}(x)))}{\tht(\uz_-(\hat {\ul \nu}(x)))}
e^{-2i(x-x_0)e_0},
\lb{350} \vspace{2mm}\\
&&q(x)=q(x_0)
\f{\tht(\uz_+(\hat {\ul \mu}(x_0)))}{\tht(\uz_-(\hat {\ul
\mu}(x_0)))}
\f{\tht(\uz_-(\hat {\ul \mu}(x)))}{\tht(\uz_+(\hat {\ul
\mu}(x)))}
e^{2i(x-x_0)e_0},
\lb{351}\vspace{2mm} \\
&&p(x_0)q(x_0)=\f{4}{\ome_0^2}
\f{\tht(\uz_+(\hat {\ul \nu}(x_0)))}{\tht(\uz_-(\hat {\ul
\nu}(x_0)))}
\f{\tht(\uz_-(\hat {\ul \mu}(x_0)))}{\tht(\uz_+(\hat {\ul
\mu}(x_0)))},
\lb{352}
\ea
and
\ba
&& {\ul \al}_{P_0}(\calD_{\hat {\ul \mu}(x)})=
  {\ul \al}_{P_0}(\calD_{\hat {\ul \mu}(x_0)})-i(x-x_0)
   \ul U_0^{(2)},
 \lb{353} \vspace{2mm} \\
&& {\ul \al}_{P_0}(\calD_{\hat {\ul \nu}(x)})=
  {\ul \al}_{P_0}(\calD_{\hat {\ul \nu}(x_0)})-i(x-x_0)
   \ul U_0^{(2)}.
 \lb{354}
\ea
\end{thm}
\begin{proof}
Since $\phi(P,x)e^{-\int_{P_0}^P \ome_{\infpm}^{(3)}}$ is
meromorphic on $\calK_n$ with divisor $\calD_{\hat {\ul \nu}(x)}-
\calD_{\hat {\ul \mu}(x)},\; \;\calD_{\hat {\ul \mu}(x)}$ is
nonspecial if and only if $\calD_{\hat {\ul \nu}(x)}$ is
nonspecial (cf.\ the comment following (\ref{a40})).
Combining (\ref{37}), (\ref{38}), (\ref{310}), (\ref{311}),
(\ref{312}), (\ref{320}), and (\ref{325}) yields the asymptotic
behavior
\ba
&& \phi(P,x)\underset{\zeta \to 0}{=}
\left\{ \begin{array}{ll}
\f{i}{2}p(x)\zeta+O(\zeta^2),
&P \mbox{ near } \inf_+, \\ \\
\f{2i}{q(x)}\zeta^{-1}+O(1),
&P \mbox{ near } \inf_-,
\end{array} \right.
\label{355} \\ \no \\
&& \psi_1(P,x,x_0)\underset{\zeta \to 0}{=}
\left\{ \begin{array}{ll}
e^{-i(x-x_0)\zeta^{-1}+O(\zeta)},
&P \mbox{ near } \inf_+, \\ \\
\left[\f{q(x)}{q(x_0)}+O(\zeta)\right]
e^{i(x-x_0)\zeta^{-1}+O(\zeta)},
&P \mbox{ near } \inf_-,
\end{array} \right.
\label{356} \\ \no \\
&& \psi_2(P,x,x_0)\underset{\zeta \to 0}{=}
\left\{ \begin{array}{ll}
\left[\f{i}{2}p(x)\zeta+O(\zeta^2\right]
e^{-i(x-x_0)\zeta^{-1}+O(\zeta)},
&P \mbox{ near } \inf_+, \\ \\
\left[\f{2i}{q(x_0)}\zeta^{-1}+O(1)\right]
e^{i(x-x_0)\zeta^{-1}+O(\zeta)},
&P \mbox{ near } \inf_-.
\end{array} \right.
\label{357}
\ea
The asymptotic behavior (\ref{356}) (near $\inf_+$), (\ref{325})
 and (\ref{343}) together with Lemma \ref{l34}
 then directly yield the theta function
representation (\ref{348}) for $\psi_1$. (\ref{343}) also
immediately yields that $\phi(P,x)$ equals (\ref{347}) which,
together with (\ref{355}), implies
\ba
&& p(x)=\f{2C(x)}{i\ome_0}
  \f{\tht(\uz_+(\hat {\ul \nu}(x)))}
  {\tht(\uz_+(\hat {\ul \mu}(x)))},
  \label{358} \\
&& q(x)=\f{2i}{C(x)\ome_0}
  \f{\tht(\uz_-(\hat {\ul \mu}(x)))}
  {\tht(\uz_-(\hat {\ul \nu}(x)))}.
  \label{359}
\ea
On the other hand (\ref{356}) (near $\inf_-$) and (\ref{348})
yield (\ref{351}). A comparison of  (\ref{351}) and (\ref{359})
determines $C(x)$ and $p(x)$ as in (\ref{350}), (\ref{352}).
Given $C(x)$, one determines $\phi$ in (\ref{347}) from
(\ref{344}) and hence $\psi_2$ as in (\ref{349}) from
$\psi_2=\phi\psi_1$. By (\ref{328}) and a special case of
Lagrange's interpolation formula,
\begin{equation}
\sum_{j=1}^n\mu_j^{k-1}
\prod_{\stackrel{\scriptstyle{\ell=1}}
{\scriptstyle{\ell \not=j}}}^{n}
(\mu_{j} - \mu_{\ell})^{-1}=
\del_{k,n},\; \; \mu_j\in\bbC,\; \;
1\leq j,k \leq n,
 \label{360}
\end{equation}
one infers
\ba
\f{d}{dx}{\ul \al}_{P_0}({\hat {\ul \mu}(x)}))=
-2i\ul c_n,\; \; \ul c_n=(c_{1,n},\ldots,c_{n,n}),
\label{361}
\ea
where
\ba
\ome_j=\sum_{k=1}^nc_{j,k}\f{\ti {\pi}^{k-1} d\ti \pi}{y}
,\; \;
1\leq j\leq n
 \label{362}
\ea
abbreviates the basis of holomorphic differentials on
$\calK_n$. By (\ref{a35}) this yields
\ba
\f{d}{dx}{\ul \al}_{P_0}({\hat {\ul \mu}(x)}))=
-i\ul U_0^{(2)}
\label{363}
\ea
and hence (\ref{353}) (respectively, (\ref{354})).
\end{proof}

For completeness we also mention another theta function
representation for the product $p(x)q(x)$, originally due to
\cite{its1}.

\begin{cor}\lb{c36}
Assume the hypotheses of Theorem \ref{t35}. Then
\begin{equation}
p(x)q(x)=-e_{0,1} -
\f{d^2}{dx^2}\ln(\tht(\uz_+(\hat {\ul \mu}(x)))).
 \lb{363a}
\end{equation}
\end{cor}
\begin{proof}
Eliminating $\psi_2(z,x)$ in (\ref{218}) results in
\begin{equation}
\psi_{1,xx}(z,x)=\f{q_x(x)}{q(x)}\psi_{1,x}(z,x) + (p(x)q(x) +
iz\f{q_x(x)}{q(x)} -z^2)\psi_1(z,x). \lb{363b}
\end{equation}
Next, using
\begin{equation}
\psi_1(P,x,x_0)\underset{\zeta \to 0}{=}
e^{-i(x-x_0)(\zeta^{-1}+e_{0,1}\zeta+O(\zeta^2))}
(1+c_1(x)\zeta + c_2(x)\zeta^2 + O(\zeta^3)), \lb{363c}
\end{equation}
one infers
\begin{align}
0  = & -\psi_{1,xx}(P,x,x_0) + \f{q_x(x)}{q(x)}\psi_{1,x}(P,x,x_0)
+ (p(x)q(x)  + i\f{q_x(x)}{q(x)}\zeta^{-1}
-\zeta^{-2})\psi_1(P,x,x_0)  \no \\
 \underset{\zeta \to 0}{=} & e^{-i(x-x_0)(\zeta^{-1}
+e_{0,1}\zeta+O(\zeta^2))}(e_{0,1} + 2ic_{1,x}(x) + p(x)q(x)
+O(\zeta)). \lb{363d}
\end{align}
By the uniqueness of $\psi_1(P,x,x_0)$ as discussed in
Lemma \ref{l34} one concludes
\begin{equation}
p(x)q(x)=-e_{0,1} - 2ic_{1,x}(x). \lb{363e}
\end{equation}
It remains to determine $c_{1,x}(x)$. First we recall from
(\ref{a24}) that
\begin{equation}
\ul \omega \underset{\zeta \to 0}{=} (\ul{c}(n) + O(\zeta))d\zeta
\,\,\, \text{near} \, \, \, \infty_+ \lb{363f}
\end{equation}
and hence
\begin{equation}
\ul{A}_{P_0}(P) \underset{\zeta \to 0}{=} \ul{A}_{P_0}(\infty_+)
+ \ul {c}(n)\zeta + O(\zeta^2)
\underset{\zeta \to 0}{=} \ul{A}_{P_0}(\infty_+)
+\f{1}{2}\ul{U}^{(2)}_0 \zeta + O(\zeta^2), \lb{363g}
\end{equation}
where we combined (\ref{341}) and (\ref{a35}) in the second equality.
Since $p(x)q(x)$ only depends on $c_{1,x}(x)$ as opposed to
$c_1(x)$ itself, it suffices to consider the following expansion near
$\infty_+$.
\begin{align}
\f{\tht(\uz(P,\hat {\ul \mu}(x)))}{\tht(\uz_+(\hat {\ul \mu}(x)))}
\underset{\zeta \to 0}{=} &1
-\f{1}{2} \f{\Sigma^n_{j=1} U^{(2)}_{0,j} \f{d}{dw_j}
\theta(\uz_+(\hat {\ul \mu}(x))+\ul w)|_{\ul w =0}}
{\tht(\uz_+(\hat {\ul \mu}(x)))}\zeta +O(\zeta^2) \no \\
=&1 + \f{1}{2i} \f{d}{dx}\ln(\tht(\uz_+(\hat {\ul \mu}(x))))\zeta
+O(\zeta^2).
\lb{363h}
\end{align}
Here we used (\ref{353}) to arrive at the last equality
in (\ref{363h}).
A comparison of (\ref{348}), (\ref{363d}), and (\ref{363h})
then yields
\begin{equation}
c_{1,x}(x) = -\f{i}{2} \f{d^2}{dx^2}
\ln(\tht(\uz_+(\hat {\ul \mu}(x)))), \lb{363i}
\end{equation}
which finally yields (\ref{363a}) employing (\ref{363e}).
\end{proof}

We note that the free constant $q(x_0)$ in (\ref{350}) (and in
(\ref{351}) using (\ref{352})) cannot be determined by this
 formalism since the AKNS equations (\ref{224}) are invariant
with respect to scale transformations. More precisely, one
has
\begin{lem}\lb{l36}
Suppose $(p,q)$ satisfies one of the AKNS equations (\ref{224})
for some $n\in\bbN_0$,
\ba
\text{AKNS}_n(p,q)=0.
\label{364}
\ea
Consider the scale transformation
\ba
(p(x,t_n),q(x,t_n))\to ({\breve p}(x,t_n),{\breve q}(x,t_n))=
(A p(x,t_n),A^{-1}q(x,t_n)),\; \; A\in \bbC\backslash
\{0\}.
\label{365}
\ea
Then
\ba
\text{AKNS}_n({\breve p},{\breve q})=0.
\label{366}
\ea
\end{lem}
\begin{proof}
Let $(D,E_{n+1})$ and $({\breve D},{\breve E}_{n+1})$ be
associated with
$(p,q)$ and $({\breve p},{\breve q})$, respectively and defined
according to
(\ref{21}) and (\ref{24}). Defining the matrix $T$ in $\bbC^2$ by
\ba
T=
\begin{pmatrix}
(A^{\f12})^{-1} & 0 \\ 0 & A^{\f12}
\end{pmatrix}
\label{367}
\ea
(fixing a particular square root branch $A^{\f12}$) one computes
\ba
&&TDT^{-1}={\breve D},
\label{368}\\
&&TE_{n+1}T^{-1}=i\sum_{\ell=0}^{n+1}
\begin{pmatrix}
-g_{n+1-\ell} & A^{-1} f_{n-\ell} \\ -Ah_{n-\ell}
& g_{n+1-\ell}
\end{pmatrix}
{\breve D}^\ell={\breve E}_{n+1}.
\label{369}
\ea
A comparison of (\ref{369}) with
\ba
{\breve E}_{n+1}=i\sum_{\ell=0}^{n+1}
\begin{pmatrix}
-{\breve g}_{n+1-\ell} & {\breve f}_{n-\ell} \\ -{\breve h}_{n-\ell}
& {\breve g}_{n+1-\ell}
\end{pmatrix}{\breve D}^\ell
\label{370}
\ea
yields
\ba
{\breve f}_{n-\ell}=A^{-1} f_{n-\ell},\; \;
{\breve g}_{n+1-\ell}= g_{n+1-\ell},\; \;
{\breve h}_{n-\ell}=A^{-1} h_{n-\ell},\; \;
0\leq \ell \leq n+1
\label{371}
\ea
and hence (\ref{366}), taking into account (\ref{224}) and
(\ref{365}).
\end{proof}

In the particular case of the nonlinear Schr\"{o}dinger (NS)
hierarchy, where
\ba
p(x,t_n)=\pm \overline{q(x,t_n)}, \; \; n\in\bbN_0,
\label{372}
\ea
(\ref{365}) further restricts $A$ to be unimodular, that is,
\ba
|A|=1.
\label{373}
\ea
In the special case of the modified Korteweg-de Vries (mKdV)
hierarchy, where
\ba
p(x,t_n)=\pm {q(x,t_n)}, \; \; n\in 2\bbN_0, \,\, c_{2\ell+1}=0, \,
\ell \in \bbN_0,
\label{374}
\ea
(\ref{365}) implies the additional restriction
\ba
A \in \{-1, +1 \}.
\label{375}
\ea

Next, we briefly consider the trivial case $n=0$ excluded in
Theorem \ref{t35}.

\begin{exmp}\lb{e37}
Assume $n=0$. Then \vspace{2mm} \\
$\calF_0(z,y)=y^2-R_2(z)=y^2-\prod_{m=0}^1(z-E_m),$\vspace{2mm} \\
$c_1=-(E_0+E_1)/2$,\vspace{2mm} \\
$p(x)=p(x_0)\exp[-2ic_1(x-x_0)],\; \;
q(x)=q(x_0)\exp[2ic_1(x-x_0)]$, \vspace{2mm} \\
$p(x)q(x)={(E_0-E_1)^2}/4$, \vspace{2mm} \\
$\phi(P,x)=\f{y(P)+z+c_1}{-iq(x)}=\f{ip(x)}{y(P)-z-c_1}$,
\vspace{2mm} \\
$\psi_1(P,x,x_0)=\exp[i(x-x_0)(y(P)+c_1)]$, \vspace{2mm} \\
$\psi_2(P,x,x_0)=\f{y(P)+z+c_1}{-iq(x_0)}\exp[i(x-x_0)(y(P)-c_1)]$.

\end{exmp}

Finally, we mention an interesting characterization of all
algebro-geometric AKNS potentials due to De Concini and Johnson
\cite{dj} in the special case where $D$ generates a self-adjoint
operator in $L^2(\bbR)\otimes \bbC^2$. In this case the
algebro-geometric potentials are characterized by the fact
that the corresponding spectrum consists of finitely many
intervals and the Lyapunov exponent vanishes a.e. on the spectrum.
In
this context we might also point out that a detailed study
of Floquet theory for periodic and self-adjoint AKNS operators
(not necessarily of algebro-geometric type) generated by $D$
can be found in \cite{grgu}.

\section{The Time-Dependent AKNS Formalism} \label{s4}

In our final section we indicate how to generalize the polynomial
approach of Sections 2 and 3 to the time-dependent AKNS hierarchy.
\newline
Our starting point is a stationary $n$-gap solution
$(p^{(0)}(x),q^{(0)}(x))$,
 associated with $\calK_n$,
\begin{align}
& i\f{p^{(0)}_x(x)}{p^{(0)}(x)} = 2c_1+ 2\sum_{j_1=1}^{ n}
\nu^{(0)}_{j_1}(x), \no\\
& \label{41}\\
& i\f{q^{(0)}_x(x)}{q^{(0)}(x)} = -2c_1- 2\sum_{j_1=1}^{ n}
\mu^{(0)}_{j_1}(x), \no
\end{align}
satisfying
\ba
\mb{AKNS}_n(p^{(0)},q^{(0)})=\left\{ \bma{l}-2{h}_{n+1}=0 \\
 -2{f}_{n+1} =0 \ema \right.
\label{42}
\ea
for some fixed $n \in \bbN_0$ and a given set of integration
constants
$\{c_{\ell}\}_{1 \leq \ell \leq n+1} $. Our principal aim is
to construct
the $r^{\text{th}}$ AKNS flow
\ba
\mb{AKNS}_r(p,q)&=&\left\{ \bma{l}p_{t_r}-2\ti{h}_{r+1}=0\\ \no
 q_{t_r}-2\ti{f}_{r+1} =0 \ema \right.  \no \\
 \label{43} \\
&=& \left\{ \bma{l} p_{t_r}+
i\ti{H}_{r,x}+2z\ti{H}_{r}-2ip\ti{G}_{r+1}=0, \no \\
    q_{t_r}-i\ti{F}_{r,x}+2z\ti{F}_{r}+2iq\ti{G}_{r+1}=0,
    \ema \right.  \no
\ea
\ba
(p(x,t_{0,r}),q(x,t_{0,r}))=(p^{(0)}(x),q^{(0)}(x)),
     \; \;x \in \bbR \quad \no
\ea
 for $t_{0,r} \in \bbR$ and some fixed
$r \in \bbN_0$. In terms of Lax pairs this amounts to solving
\ba
&&\f{d}{dt_r}D(t_r)-[\ti{E}_{r+1}(t_r),D(t_r)]=0,\; \;
t_r \in \bbR,
 \label{44} \\
&& [{E}_{n+1}(t_{0,r}),D(t_{0,r})]=0. \label{44a}
\ea
As a consequence one obtains that
\ba
&&[E_{n+1}(t_r),D(t_r)]=0,\; \; t_r \in \bbR, \label{45} \\
&&E_{n+1}(t_r)^2= -R_{2n+2}(D(t_r))= -\prod_{m=0}^ {2n+1}
(D(t_r)-E_m),\; \;
t_r \in \bbR
\label{46}
\ea
since the \mb{AKNS} flows are isospectral deformations of
$D(t_{0,r})$.

We emphasize that the integration constants $\{\ti c_{\ell}\}$
in $\ti{E}_{r+1}$
and $\{c_{\ell}\}$ in $E_{n+1}$, in general, are independent
of each other (even
 if $r=n$). Hence we shall
employ the notation
$\ti{E}_{r+1},\; \;\ti{V}_{r+1},\; \; \ti{F}_{r},\;
\;\ti{G}_{r+1}$
$\ti{H}_{r},\; \;\ti{f}_{l},\; \;\ti{g}_{l},
\; \;\ti{h}_{l},\; \;\ti{c}_{l},$ etc.,
 in order
to distinguish it from
$E_{n+1},\; \;V_{n+1},\; \; F_{n},\; \;G_{n+1},\; \;H_{n},\;
\;f_{l},$
$g_{l},\; \;h_{l},\; \;c_{l},$ etc. In addition,
we followed a more elaborate notation inspired by Hirota's
$\tau$- function
 approach
and indicated the individual $r^{\text{th}}$ \mb{AKNS} flow by
a separate time
 variable
$t_r \in \bbR$. (The latter notation suggests considering all
\mb{AKNS} flows
simultaneously by introducing $\ul{t}=(t_0,t_1,t_2,........)$.)

Instead of working directly with (\ref{44}), (\ref{44a}) and
(\ref{45}),
it is more convenient to
take the zero-curvature equations (\ref{233}) as our point of
departure,
that is, we start from
\ba
U_{t_{r}}-\ti{V}_{r+1,x} +[U,\ti{V}_{r+1}]=0,\; \; (x,t_r)
\in \bbR^2,
\label{47}
\ea
\ba
-V_{n+1,x} +[U,V_{n+1}]=0,\; \; (x,t_r) \in \bbR^2,
\label{48}
\ea
where (cf. (\ref{29}))
\ba
&& \quad U(z,x,t_r) = \begin{pmatrix}
                  -iz & q(x,t_r) \\ p(x,t_r) & iz
                   \end{pmatrix}, \no \\
&& \quad  \ti{V}_{r+1}(z,x,t_r) = i \begin{pmatrix}
                  -\ti{G}_{r+1}(z,x,t_r) & \ti{F}_r(z,x,t_r) \\
           -\ti{H}_r(z,x,t_r) &
                    \ti{G}_{r+1}(z,x,t_r)
                   \end{pmatrix}, \label{49}\\
&& \quad  V_{n+1}(z,x,t_r) = i \begin{pmatrix}
                  -G_{n+1}(z,x,t_r)&F_n(z,x,t_r)\\-H_n(z,x,t_r)&
                    G_{n+1}(z,x,t_r)
                   \end{pmatrix}, \no
\ea
\ba
&&F_n(z,x,t_r)= \sum_{{\ell}=0}^{n}f_{n-{\ell}}(x,t_r)z^{\ell},
      \; \; f_0(x,t_r)=-iq(x,t_r), \no \\ \label{410} \\
&&F_n(z,x,t_{0,r})=F^{(0)}_n(z,x)
=\sum_{{\ell}=0}^{n}f^{(0)}_{n-{\ell}}(x)
       z^{\ell},
      \; \; f^{(0)}_0(x)=-iq^{(0)}(x), \no \\
&&G_{n+1}(z,x,t_r)=
\sum_{{\ell}=0}^{n+1}g_{n+1-{\ell}}(x,t_r)z^{\ell},
      \; \; g_0(x,t_r)=1, \no \\ \label{411} \\
&&G_{n+1}(z,x,t_{0,r})=G^{(0)}_{n+1}(z,x)=
\sum_{{\ell}=0}^{n+1}g^{(0)}_
 {n+1-{\ell}}(x)z^{j},
      \; \; g^{(0)}_0(x)=1, \no \\
&&H_n(z,x,t_r)= \sum_{{\ell}=0}^{n}h_{n-{\ell}}(x,t_r)z^{\ell},
     \; \; h_0(x,t_r)=ip(x,t_r), \no \\ \label{412}\\
&&H_n(z,x,t_{0,r})= H^{(0)}_n(z,x)=
\sum_{{\ell}=0}^{n}h^{(0)}_{n-{\ell}}
     (x)z^{\ell},
     \; \; h^{(0)}_0(x)=ip^{(0)}(x) \no
\ea
for fixed $t_{0,r} \in \bbR, \; n \in \bbN_0,\; r \in \bbN_0$.
 Here $f_{\ell}(x,t_r)$ and $f^{(0)}_{\ell}(x)$ are defined as
in (\ref{22})
 with
$(p(x),q(x))$ replaced by $(p(x,t_r),q(x,t_r))$ and
$(p^{(0)}(x),q^{(0)}(x))$,
respectively.  Explicitly, (\ref{47}) and (\ref{48}) are
equivalent to
\ba
&& p_{t_r}=-i\ti{H}_{r,x}-2z\ti{H}_r+2ip\ti G_{r+1}, \label{412a}\\
 && q_{t_r}= i\ti F_{r,x}-2z\ti F_r-2iq\ti G_{r+1}, \label{412b} \\
 && \ti{G}_{r+1,x}= p\ti{F}_r + q \ti{H}_r \label{412c}
\ea
and (cf. (\ref{210})--(\ref{212}))
\ba
&&F_ {n,x} =  -2 i z F_n +2 q G_{n+1}, \label{412d}\\
&&G_ {n+1,x}  =  pF_n+q  H_n, \label{412e} \\
&&H_ {n,x}  =  2 i z H_n  +2 p G_{n+1} , \label{412f}
\ea
respectively. In particular, (\ref{214}) holds in the present
$t_r$-dependent setting, that is,
\ba
G^{2}_{n+1}-F_nH_n=R_{2n+2}.
\label{412g}
\ea
In analogy to
 (\ref{311}) and (\ref{312}) we write
\ba
F_n(z,x,t_r)= -iq(x,t_r)\prod_{j=1}^{n} (z-\mu_j(x,t_r)),
\label{413}
\ea
\ba
H_n(z,x,t_r)= ip(x,t_r)\prod_{j=1}^{n} (z-\nu_j(x,t_r))
\label{414}
\ea
and define in analogy to (\ref{310}), the following meromorphic
function
$\phi(P,x,t_r)$ on $\calK_n$, the fundamental ingredient for
constructing
algebro-geometric solutions of the time-dependent \mb{AKNS}
hierarchy,
\ba
\phi(P,x,t_r)=\f{y(P)+G_{n+1}(z,x,t_r)}{F_n(z,x,t_r)}=
     \f{-H_n(z,x,t_r)}{y(P)-G_{n+1}(z,x,t_r)},  \label{415}\\
\; \;\; \; \qquad\qquad\qquad P=(z,y) \in \calK_n. \no
\ea
As in (\ref{313}) and (\ref{314}) one introduces
\begin{align}
\hat{\mu}_j (x,t_r)
&=({\mu_j }(x,t_r),G_{n+1}({\mu_j }(x,t_r),x,t_r))\in \calK_n,
\, \, \,  1 \leq j \leq n, \; \; (x,t_r)\in \bbR^2, \label{416}\\
\hat{\nu}_j (x,t_r)
&=({\nu_j }(x,t_r),-G_{n+1}({\nu_j }(x,t_r),x,t_r))\in \calK_n,
\, \, \,  1 \leq j \leq n, \; \; (x,t_r)\in \bbR^2, \label{417}
\end{align}
and infers that the divisor $(\phi(P,x,t_r))$ of $\phi(P,x,t_r)$ is
given by
\ba
\left(\phi(P,x,t_r)\right)={\calD}_{\hat{\underline \nu}(x,t_r)}(P) -
{\calD}_{\hat{\underline \mu}(x,t_r)}(P)+{\calD}_{\infty_{+}}(P) -
{\calD}_{\infty_{-}}(P).
\label{418}
\ea

Next we define the time-dependent BA-function
$\Psi(P,x,x_0,t_r,t_{0,r})$
by
\ba
\Psi(P,x,x_0,t_r,t_{0,r})=\begin{pmatrix}
             \psi_1(P,x,x_0,t_r,t_{0,r}) \\
              \psi_2(P,x,x_0,t_r,t_{0,r})
     \end{pmatrix},
\label{421}
\ea
\ba
&&\psi_1(P,x,x_0, t_r,t_{0,r}) =
\exp \left\{ \int_{x_0}^x \, dx' [-iz+
q(x',t_r)\phi(P,x',t_r)]  \right. \no \\
&&\qquad \qquad \qquad \qquad
 \left. +i\int_{t_{0,r}}^{t_r} \, ds [\ti{F}_r (z,x_0,
s) \phi (P,x_0,s) -\ti{G}_{r+1} (z,x_0, s)] \right\},
 \label{422} \\
&&\psi_2(P,x,x_0,t_r,t_{0,r})=\phi(P,x,t_r)\psi_1(P,x,x_0,t,t_{0,r}),
\label{423}\\
&& \quad\qquad \qquad \qquad \qquad \qquad \qquad \qquad \qquad
 P\in \calK_n\backslash \{\infty_{\pm}\}, \; \; (x,t_r)
\in \bbR^2, \no
\ea
with fixed $(x_0,t_{0,r}) \in \bbR^2$. The following Lemma records
properties of
$\phi(P,x,t_r)$ and $\Psi(P,x,x_0, t_r,t_{0,r})$ in analogy to
the stationary
 case
discussed in Lemma 3.1.
\begin{lem}\label{l41}
Assume (\ref{412a})--(\ref{412g}), $P=(z,y) \in \calK_n
\backslash \{
\infty_{\pm}\}$, and let $(z,x,x_0,t_r,t_{0,r})$
$ \in \bbC \times \bbR^4$. Then
\ba
&&(i).\; \; \phi(P,x,t_r)\; \;   \text{satisfies}
\no \vspace{2mm}\\
&& \quad\quad  \phi_{x}(P,x,t_r)+q(x,t_r)\phi(P,x,t_r)^2-
2iz\phi(P,x,t_r)
     =p(x,t_r), \label{424}\vspace{2mm} \\
&& \quad \quad  [q(x,t_r)\phi(P,x,t_r)]_{t_{r}}=
i \partial_x[\ti{F}_r(z,x,t_r)
    \phi(P,x,t_r)-\ti{G}_{r+1}(z,x,t_r)],
\label{425}\vspace{2mm} \\
&& \quad \quad  \phi_{t_{r}}(P,x,t_r)=
2i\ti{G}_{r+1}(z,x,t_r)\phi(P,x,t_r)
          +
\f{1}{q(x,t_r)}[-i\ti{G}_{r+1,x}(z,x,t_r)\no \\
&& \quad\quad \quad \quad \quad \quad +i\ti{F}_r(z,x,t_r)
    \phi_{x}(P,x,t_r) +2z\ti{F}_r(z,x,t_r)\phi(P,x,t_r)]
   .\label{426}\vspace{2mm} \\
\no \\
&&(ii). \; \; \psi_j(P,x,x_0, t_r,t_{0,r}),\; \; j=1,2\; \;
\text{satisfy}
\no \vspace{2mm} \\
&& \quad \quad \psi_{1,x}(P,x,x_0, t_r,t_{0,r})=
[q(x,t_r)\phi(P,x,t_r)-iz]
     \psi_{1}(P,x,x_0, t_r,t_{0,r}),
\label{427}\vspace{2mm}\\
&& \quad \quad  \psi_{1,t_r}(P,x,x_0, t_r,t_{0,r})=
i[\ti{F}_r(z,x,t_r)
    \phi(P,x,t_r)-\ti{G}_{r+1}(z,x,t_r)]\times
\vspace{2mm}\no\\
&&  \quad \quad \quad \quad \quad\quad \quad \quad
\quad \quad \quad \quad
  \times\psi_{1}(P,x,x_0, t_r,t_{0,r}),
    \label{428}\\
&& \quad \quad  \psi_{2,x}(P,x,x_0,t_r,t_{0,r})=
[p(x,t_r)\phi(P,x,t_r)^{-1}+iz]
     \psi_{1}(P,x,x_0, t_r,t_{0,r}),
\label{429}\vspace{2mm}\\
&& \quad \quad  \psi_{2,t_r}(P,x,x_0, t_r,t_{0,r})=
-i[\ti{H}_r(z,x,t_r)
    \phi(P,x,t_r)^{-1}-\ti{G}_{r+1}(z,x,t_r)]\times
\vspace{2mm}\no\\
&& \quad \quad \quad \quad  \quad\quad \quad \quad
\quad \quad \quad \quad
    \times\psi_{2}(P,x,x_0, t_r,t_{0,r}),
\label{430} \\
&& \mb{or equivalently,} \no \\ \no \\
&& \quad \quad  \Psi_x(P,x,x_0, t_r,t_{0,r})=
U(z,x,t_r)\Psi(P,x,x_0,
 t_r,t_{0,r}),  \no \\
&& \quad \quad  iy(P)\Psi(P,x,x_0, t_r,t_{0,r})=
V_{n+1}(z,x,t_r)\Psi(P,x,x_0,
 t_r,t_{0,r}), \label{431}\\
&& \quad \quad  \Psi_{t_r}(P,x,x_0, t_r,t_{0,r})=
\tilde V_{r+1}(z,x,t_r)\Psi(P,x,x_0,
 t_r,t_{0,r})
 \no
\ea

({i.e}., $(D-z)\Psi={0}, \; \; (E_{n+1}-iy)\Psi={0},\;
\;\Psi_{t_r}=
\ti{E}_{r+1}\Psi).$
\ba
&&(iii). \; \; \phi(P,x,t_r)\phi(P^*,x,t_r)=
  \f{H_{n} (z, x,t_r)}{F_{n} (z, x,t_r)}.
 \quad  \quad  \quad  \quad  \quad  \quad  \quad
\quad  \quad  \quad  \label{432} \vspace{2mm}\\
&&(iv). \; \; \phi(P,x,t_r)+\phi(P^*,x,t_r)=
    \f{2G_{n+1} (z, x,t_r)}{F_{n} (z, x,t_r)}.
\quad  \quad  \quad  \quad  \quad  \quad  \quad
\quad  \quad  \quad \label{433} \vspace{2mm}\\
&&(v). \; \; \phi(P,x,t_r)-\phi(P^*,x,t_r)= \f{2 y(P)}
    {F_{n} (z, x)}.
\quad  \quad  \quad  \quad  \quad  \quad  \quad
\quad  \quad  \quad \label{434}
\ea
\end{lem}
 \begin{proof}
(\ref{424}) follows from (\ref{48}) and (\ref{415}).
(\ref{425}) can
be proven as follows. Using (\ref{43}) and (\ref{424})
one infers by
a straight forward (but rather lengthy) calculation that
\ba
(\partial_x+2q\phi-2iz-\f{q_x}{q})[(q\phi)_t-
i(\ti{F}_r\phi-\ti{G}_{r+1})_x]
=0.
\label{435}
\ea
Thus
\ba
(q\phi)_t-i(\ti{F}_r\phi-\ti{G}_{r+1})_x=
C e^{\int^{x}dx'[2iz+\f{q_{x'}}{q}-2q\phi] },
\label{436}
\ea
where $C$ is independent of $x$ (but may depend on $P$ and
$t_r$).
By inspection of (\ref{415}), the left-hand side of
(\ref{436}) is
meromorphic on $\calK_n$ while the right-hand side of
(\ref{436}) is not
meromorphic at $\infty_{+}$ and $\infty_{-}$ unless $C=0$.
Hence one infers
$C=0$ and thus (\ref{425}). (\ref{426}) is then an immediate
consequence of
(\ref{43}) (i.e., the \mb{AKNS} equation for $q_{t_r}$) and
(\ref{425}).
(\ref{427}) is clear from (\ref{422}) and (\ref{429}) is obvious
from (\ref{423}), (\ref{424}), and (\ref{427}).
(\ref{428}) follows from (\ref{422}), and
(\ref{430}) is a straightforward consequence of (\ref{423}),
 and (\ref{428}). Finally, $(iii)$--$(v)$ are proved as in
Lemma \ref{l31}.
\end{proof}

Next we consider the $t_r$-dependence of $F_n(z,x,t_r),\;
\;G_{n+1}(z,x,t_r)$,
 and
$H_n(z,x,t_r)$.

\begin{lem} \label{l42}
Assume (\ref{412a})--(\ref{412g}) and let $(z,x,t_r)\in
\bbC \times \bbR^2.$
 Then
\ba
&&(i). \; \; F_{n,t_{r}}(z,x,t_r)=
2i[\ti{F}_r(z,x,t_r)G_{n+1}(z,x,t_r)-
         F_n(z,x,t_r)\ti{G}_{r+1}(z,x,t_r)]. \label{437}\\
&&(ii).\; \;  G_{n+1,t_{r}}(z,x,t_r)=
i[\ti{F}_r(z,x,t_r)H_{n}(z,x,t_r)-
         F_n(z,x,t_r)\ti{H}_{r}(z,x,t_r)].\; \; \; \; \quad
\quad \quad \quad
 \label{438}\\
&&(iii). \;H_{n,t_{r}}(z,x,t_r)=
-2i[\ti{H}_r(z,x,t_r)G_{n+1}(z,x,t_r)-
         H_n(z,x,t_r)\ti{G}_{r+1}(z,x,t_r)].  \label{439}
\ea
In particular, (i)--(iii) are equivalent to
\begin{equation}
-V_{n+1,t_{r}} + [\ti V_{r+1} , V_{n+1}] =0.
\label{438a}
\end{equation}
\end{lem}

\begin{proof}
By (\ref{415}), (\ref{426}), (\ref{433}), and (\ref{434})
one infers
\ba
\phi_{t_r}(P)- \phi_{t_r}(P^*)=-\f{2 y(P)F_{n,t_r}}{F_n^2}=
    \f{4iy(P)}{F_n^2}(\ti{G}_{r+1}F_n - \ti{F}_rG_{n+1}),
\label{440}
\ea
which proves (\ref{437}). Similarly, differentiating (\ref{433})
 with
respect to $t_r$, using (\ref{424}), (\ref{426}),
(\ref{432})--(\ref{434}),
 and (\ref{412c}), proves (\ref{438}).\;(\ref{439}) finally
follows from
$(G_{n+1}^2 - F_nH_n)_{t_{r}}=0$ (cf. (\ref{412g})), (\ref{437}),
and
(\ref{438}).
\end{proof}

The remaining items $(vi)$--$(ix)$ of Lemma \ref{l31} in the
present
time-dependent setting then read

\begin{lem} \label{l43}
Assume (\ref{412a})--(\ref{412g}),
$P=(z,y) \in \calK_n \backslash \{
\infty \},$ and let $(z,x,x_0,t_r,t_{0,r})$
$\in \bbC \times \bbR^4.$ Then
\ba
&&(i). \; \; {\psi}_{1}(P,x,x_0,t_r,t_{0,r}){\psi}_{1}
(P^*,x,x_0,t_r,t_{0,r})=
    \f{F_{n} (z, x,t_r)}{F_{n} (z, x_0,t_{0,r})}. \label{441} \\
&&(ii). \; \; {\psi}_{2}(P,x,x_0,t_r,t_{0,r})
{\psi}_{2}(P^*,x,x_0,t_r,t_{0,r})
   =\f{H_{n} (z, x,t_r)}{F_{n} (z, x_0,t_{0,r})}. \label{442} \\
&&(iii). \; \; {\psi}_{1}(P,x,x_0,t_r,t_{0,r})=
\left[\f{F_{n} (z, x,t_r)}
     {F_{n} (z, x_0,t_{0,r})} \right]^{\f12} \times \no \\
&& \quad\quad\quad\quad\quad \times\exp \left\{y(P)
   \left( \int_{x_{0}}^{x} dx'\f{q(x',t_r)}{F_n(z,x',t_r)}
+ \int_{t_{0,r}}^{t_r} ds\f{\ti{F}_r(z,x_0,s)}{F_n(z,x_0,s)}
\right) \right\}. \label{443} \\
&&(iv).\; \;{\psi}_{1}(P,x,x_0,t_r,t_{0,r})
{\psi}_{2}(P^*,x,x_0,t_r,t_{0,r}) +
      {\psi}_{1}(P^*,x,x_0,t_r,t_{0,r}){\psi}_{2}
(P,x,x_0,t_r,t_{0,r}) \no \\
&& \quad \quad\quad\quad\quad  = \f{2 G_{n+1} (z, x,t_r)}
{F_{n} (z,
 x_0,t_{0,r})}.
 \label{444}
\ea
\end{lem}

\begin{proof}
(\ref{441}) follows from (\ref{412d}), (\ref{422}), (\ref{433}),
and
(\ref{437}). (\ref{442}) is clear from (\ref{423}), (\ref{432}),
and
(\ref{441}).\ (\ref{443}) follows from (\ref{412d}), (\ref{422}),
(\ref{433}), (\ref{434}), and (\ref{437}) by splitting
$\phi(P)= \f12[\phi(P)+\phi(P^*)] + \f12[\phi(P)-\phi(P^*)]$ in
(\ref{422}).
Finally, (\ref{444}) is clear from (\ref{423}), (\ref{433}), and
(\ref{441}).
\end{proof}

The dynamics of the zeros $\mu_j(x,t_r)$ and $\nu_j(x,t_r)$ of
$F_n(z, x,t_r)$
and $H_n(z, x,t_r)$, in analogy to Lemma \ref{l32}, is then
described in

\begin{lem} \label{l44}
Assume (\ref{412a})--(\ref{414}) and let $(x,t_r)\in \bbR^2.$
Then
\ba
&& (i).\quad \mu_{j,x}(x,t_r)  = \f{-2iy({\hat{\mu}_j  }(x,t_r))}
{\prod_{\stackrel{\scriptstyle{k=1}}{\scriptstyle{k \not=j}}}^{n}
(\mu_{j}(x,t_r) - \mu_{k}(x,t_r))},
\quad 1 \leq j \leq n, \label{445} \\
&& \quad \mu_{j,t_{r}}(x,t_r) = \f{2\ti{F}_r(\mu_j (x,t_r),x,t_r)y
(\hat{\mu}_j  (x,t_r)) }
{q(x,t_r)\prod_{\stackrel{\scriptstyle{k=1}}
{\scriptstyle{k \not=j}}}^{n}
(\mu_{j}(x,t_r) - \mu_{k}(x,t_r))},
\quad 1 \leq j \leq n. \label{446} \\
&& (ii).\quad \nu_{j,x}(x,t_r) = \f{-2iy(\hat{\nu}_j  (x,t_r))}
{\prod_{\stackrel{\scriptstyle{k=1}}{\scriptstyle{k \not=j}}}^{n}
(\nu_{j}(x,t_r) - \nu_{k}(x,t_r))},
\quad 1 \leq j \leq n,
\label{447}\\
&& \quad \nu_{j,t_{r}}(x,t_r)  =
\f{-2\ti{H}_r(\nu_j (x,t_r),x,t_r))
y(\hat{\nu}_j  (x,t_r))}
{p(x,t_r)\prod_{\stackrel{\scriptstyle{k=1}}
{\scriptstyle{k \not=j}}}^{n}
(\nu_{j}(x,t_r) - \nu_{k}(x,t_r))},
\quad 1 \leq j \leq n. \label{448}
\ea
\end{lem}

\begin{proof}
(\ref{445}) and (\ref{447}) are proved as in Lemma \ref{l32}
and follow from (\ref{413}), (\ref{414}), (\ref{412d}),
(\ref{416}) and (\ref{417}). Similarly, (\ref{446}) and
(\ref{448}) follow from (\ref{413}), (\ref{416}), (\ref{437})
and (\ref{414}), (\ref{417}), (\ref{439}), respectively.
\end{proof}

The initial condition
\ba
(p(x,t_{0,r}),q(x,t_{0,r}))=(p^{(0)}(x),q^{(0)}(x)),\; \; x
\in \bbR
\lb{449}
\ea
in (\ref{43}) is taken care of by
\ba
\hat{\mu}_j  (x,t_{0,r})=\hat{\mu}_j ^{(0)} (x),\; \;
\hat{\nu}_j  (x,t_{0,r})=\hat{\nu}_j ^{(0)} (x),\; \;
1\leq j\leq n,\; \; x \in \bbR
\lb{450}
\ea
(cf. (\ref{410}), (\ref{412}) and (\ref{413}), (\ref{414})).

The trace relations in Lemma 3.3 extend in a
one-to-one manner to the present time-dependent setting
 by substituting
\ba
& (p(x),q(x)) \rightarrow (p(x,t_r),q(x,t_r)), \no
\\ \lb{451} \\
& (\mu_j(x),\nu_j(x)) \rightarrow (\mu_j(x,t_r),\nu_j(x,t_r)),
 \quad 1 \leq j \leq n, \no
\ea
keeping $\{c_\ell\}_{1 \leq \ell \leq n}$ as in (\ref{332})
since $\calK_n$ is $t_r$-dependent.

It remains to provide the explicit
theta function representation of $\Psi$, $\phi$, $p$,
and $q$. We rely on the notation established in Section 3
and Appendix A in the following, assuming $\calK_n$ to be
nonsingular as in (\ref{333}). As in Section 3 we
assume $n\in\bbN$ for the remainder of this argument.

In addition to (\ref{334})--(\ref{342})
we need to introduce the Abelian differentials
of the second kind $\ome_{\inf_{\pm},r}^{(2)}$
(cf. (\ref{a34}), (\ref{a35})) defined by
\begin{equation}
\ome_{\infty_\pm,r}^{(2)}
\underset{\zeta \to 0}{=}[\zeta^{-2-r} +O(1)]
\, d\zeta \text{ near }
\infty_\pm,\; \; r\in\bbN_0,
\lb{452}
\end{equation}
\begin{equation}
 \int_{a_j} \ome_{\inf_{\pm},r}^{(2)} = 0,
\quad 1\leq j\leq n,
\lb{453}
\end{equation}
\begin{align}
&\ti{\ul U}_r^{(2)} =(\ti U_{r,1}^{(2)},
\ldots ,\ti U_{r,n}^{(2)}),\; \;
\ti U_{r,j}^{(2)} = \f{1}{2\pi i}
\int_{b_j}\ti \Ome_{r}^{(2)},\; \;
\ti \Ome_{r}^{(2)}= \sum_{q=0}^r (q+1) \ti c_{r-q}
(\ome_{\inf_+,q}^{(2)}-\ome_{\inf_-,q}^{(2)}),
\lb{454}
\end{align}
\begin{align}
\int_{P_0}^P\ti \Ome_{r}^{(2)}
\underset{\zeta \to 0}{=}\mp \left[\sum_{q=0}^r \ti c_{r-q}
\zeta^{-1-q}+\ti e_{r,0}
+O(\zeta)\right], \; \; P=(\zeta^{-1},y)\; \;
\mb{near}\; \; \inf_{\pm},
\lb{455}
\end{align}
with $\{\ti c_\ell\}_{1\leq \ell \leq r},\; \; \ti c_0=1$
the integration constants in $\ti F_r$. Moreover, writing
\begin{equation}
\ome_j =\left( \sum_{m=0}^\infty d_{j,m}
 (\infty_\pm) \zeta^m \right)
d\zeta
=\pm \left( \sum_{m=0}^\infty d_{j,m} (\infty_+)
 \zeta^m \right)d
\zeta \text{ near } \infty_\pm,
\lb{456}
\end{equation}
relation (\ref{a35}) yields
\begin{equation}
\ti U_{r,j}^{(2)}
= 2 \sum_{q=0}^r \ti c_{r-q} d_{j,q}(\infty_+),\quad 1\leq j \leq n.
\lb{457}
\end{equation}

Before we can prove the main result of this section we need
 the following auxiliary result which is of independent
interest due to its implications for the Green's matrix of the
differential expression $D$.

\begin{lem} \lb{l45}
Let $P\in\calK_n\backslash\{\inf_+,\inf_-\}$ and $(x,t_r)
\in \bbR^2$. Then, for $P=(\zeta^{-1},y)$ near $\inf_{\pm}$,
one obtains the asymptotic expansions
\ba
\f{F_n(\zeta^{-1},x,t_r)}{y(P)}\underset{\zeta \to 0}{=}
\mp\zeta\sum_{k=0}^{\inf}\hat f_k(x,t_r)\zeta^k, \quad
\hat f_0(x,t_r)=-iq(x,t_r), \lb{4.58}
\ea
\ba
\f{H_n(\zeta^{-1},x,t_r)}{y(P)}\underset{\zeta \to 0}{=}
\mp\zeta\sum_{k=0}^{\inf}\hat h_k(x,t_r)\zeta^k, \quad
\hat h_0(x,t_r)=ip(x,t_r), \lb{4.58a}
\ea
where $\hat f_k(x,t_r)$ and $\hat h_k(x,t_r)$
denote the homogeneous
coefficients $f_k(x,t_r)$ and $ h_k(x,t_r)$
 in (\ref{410}) and (\ref{412}) (i.e., the ones
satisfying (\ref{22}) with all integration constants
$c_\ell=0,\; \ell \in \bbN$). Explicitly, $\hat f_k(x,t_r)$
and $\hat h_k(x,t_r)$
can be computed from the recursion relations,
\ba
&&\hat f_0=-iq, \quad \hat f_1=\f12q_x, \no \\
&&\hat f_k=\sum_{\ell=0}^{k-2}\left(-\f{i}{4q}\hat f_\ell
 \hat f_{k-2-\ell,xx}
+\f{iq_x}{4q^2}\hat f_{\ell}\hat f_{k-2-\ell,x}+\f{i}{8q}
\hat f_{\ell,x}\hat f_{k-2-\ell,x} \right.\lb{459} \\
&&\quad\left.+\f{ip}{2}\hat f_{\ell}\hat f_{k-2-\ell}\right)
-\f{q_x}{2q^2}\sum_{\ell=0}^{k-1}\hat f_{\ell}
\hat f_{k-1-\ell}-\f{i}{2q}
\sum_{\ell=1}^{k-1}\hat f_{\ell}\hat f_{k-\ell},
\quad k \geq 2 \no
\ea
and
\ba
&&\hat h_0=ip, \quad \hat h_1=\f12p_x, \no \\
&&\hat h_k=\sum_{\ell=0}^{k-2}\left(\f{i}{4p}\hat h_\ell
 \hat h_{k-2-\ell,xx}
-\f{ip_x}{4p^2}\hat h_{\ell}\hat h_{k-2-\ell,x}-\f{i}{8p}
\hat h_{\ell,x}\hat h_{k-2-\ell,x} \right.\lb{459a} \\
&&\quad\left.-\f{iq}{2}\hat h_{\ell}\hat h_{k-2-\ell}\right)
-\f{p_x}{2p^2}\sum_{\ell=0}^{k-1}\hat h_{\ell}
\hat h_{k-1-\ell}+\f{i}{2p}
\sum_{\ell=1}^{k-1}\hat h_{\ell}\hat h_{k-\ell},
\quad k \geq 2. \no
\ea
\end{lem}
\begin{proof}
Define
\ba
\hat F(P,x,t_r)=\f{F_n(\zeta^{-1},x,t_r)}{y(P)}
\lb{460}
\ea
then
\ba
\hat F\hat F_{xx}-\f{q_x}{q}\hat F\hat F_{x}-
\f12\hat F_{x}^2+2(\zeta^{-2}-i\zeta^{-1}\f{q_x}{q}
-pq)\hat F^2=-2q^2 \lb{461}
\ea
by (\ref{216a}). Moreover,
\ba
&&\hat F(P,x,t_r)=\mp\zeta\f{\sum_{\ell=0}^{n}
f_\ell(x,t_r)\zeta^{\ell}}
{[\prod_{m=0}^{2n+1}(1-E_m)\zeta]^{\f12}},\quad
P\in\Pi_{\pm}
\lb{462} \\
&&\quad \underset{\zeta \to 0}{=}
\mp\zeta\sum_{k=0}^{\inf}
  \hat f_k(x,t_r) \zeta^k\quad \text{near} \quad \inf_{\pm},
\quad f_0(x,t_r)=\hat f_0(x,t_r)=-iq(x,t_r),\quad
\lb{463}
\ea
where we chose the branch
\ba
\left[\prod_{m=0}^{2n+1}(1-E_m\zeta)\right]^{\f12}
\underset{\zeta \to 0}{=}1-\f12
\left(\sum_{m=0}^{2n+1}E_m\right)\zeta
+O(\zeta^2).
\lb{464}
\ea
Insertion of (\ref{463}) into (\ref{461}) yields the
recursion relation (\ref{459}) which represents the
homogeneous solutions for the $f_\ell$ due to the lack
 of any possible integration constants in (\ref{459}).
(\ref{4.58a}) and (\ref{459a}) are proved in the same manner
using (\ref{216c}).
\end{proof}

The recursion technique in
Lemma \ref{l45} represents the AKNS analog of the recursive KdV
approach in Sections~2--4 of \cite{sch}.

Lemma \ref{l45} has interesting consequences for the asymptotic
high-energy expansion of the Green's matrix $G(z,x,x')$ of $D$
(i.e., the integral kernel of the resolvent $(D-z)^{-1}$) as
described in the following remark.

\begin{rem} \lb{r46}
Let $D$ be given by (\ref{21}) ($p,\; q$ not necessarily
algebro-geometric coefficients) and assume that for some $z\in\bbC$,
and all $x_0\in \bbR$,
 $\psi_{1,\pm}(z,.),\; \psi_{2,\pm}(z,.)\in L^2((x_0,\pm\infty))$
satisfy (\ref{218}), that is,
\ba
\psi_{1,\pm,x}=-iz\psi_{1,\pm}+q\psi_{2,\pm},\;
\psi_{2,\pm,x}=iz\psi_{2,\pm}+p\psi_{1,\pm}.
\lb{473a}
\ea

Then the Green's matrix $G(z,x,x'),\; x\neq x'$ of $D$ is
given by
\ba
G(z,x,x')= \f{i}{W(z)}\left\{
\bma{l}
\begin{pmatrix}
\psi_{1,+}(z,x)\psi_{2,-}(z,x') & \psi_{1,+}(z,x)\psi_{1,-}(z,x') \\
\psi_{2,+}(z,x)\psi_{2,-}(z,x') & \psi_{2,+}(z,x)\psi_{1,-}(z,x')
\end{pmatrix},\; x>x', \\ \\
\begin{pmatrix}
\psi_{1,-}(z,x)\psi_{2,+}(z,x') & \psi_{1,-}(z,x)\psi_{1,+}(z,x') \\
\psi_{2,-}(z,x)\psi_{2,+}(z,x') & \psi_{2,-}(z,x)\psi_{1,+}(z,x')
\end{pmatrix},\; x<x',
\ema \right.
\lb{474a}
\ea
where the Wronskian
\ba
W(z):=\left| \bma{l}
\psi_{1,-}(z,x)\; \;  \psi_{1,+}(z,x) \\
\psi_{2,-}(z,x)\; \;  \psi_{2,+}(z,x)
\ema \right|
=\psi_{1,-}(z,x)\psi_{2,+}(z,x)
 -\psi_{2,-}(z,x)\psi_{1,+}(z,x)
\lb{475a}
\ea
is $x$-independent. Note that $G(z,x,x')$ is continuous
at $x=x'$ in its off-diagonal elements but discontinuous on
the diagonal.
\newline
In the special algebro-geometric context we may replace
\ba
\psi_{j,+}(z,x) \,\, \text{by} \,\,\psi_j(P,x,x_0),\;
\psi_{j,-}(z,x) \,\, \text{by} \,\,\psi_j(P^*,x,x_0),\; j=1,2,\;
P=(z,y)
\lb{476a}
\ea
and (cf. (\ref{37}), (\ref{38}), and (\ref{322}))
\ba
W(z) \,\, \text{by} \, W(P)=\left| \bma{ll}
\; \; \; \; 1& \; \; \; \;  1 \\
\phi(P^*,x_0)& \phi(P,x_0)
\ema \right|
=\phi(P,x_0)- \phi(P^*,x_0)
=\f{2y(P)}{F_n(z,x_0)}
\lb{477a}
\ea
since $W$ is $x$-independent. Substituting (\ref{476a}) and
(\ref{477a}) into (\ref{474a}), denoting the result by
$G(P,x,x')$, then yields
\ba
&&\f12[G(P,x,x+0)+G(P,x,x-0)]=\f12[G(P,x-0,x)+G(P,x+0,x)] \no \\
\lb{478a}\\
&&=\f{i}{2y(P)}
\begin{pmatrix}
G_{n+1}(z,x) & F_{n}(z,x) \\
H_{n}(z,x) & G_{n+1}(z,x)
\end{pmatrix}
=\f{i}{2}
\begin{pmatrix}
\hat G(P,x) & \hat F(P,x) \\
\hat H(P,x) & \hat G(P,x)
\end{pmatrix},\; P=(z,y),\no
\ea
where (cf. (\ref{460}))
\ba
\hat F(P,x)=\f{F_n(z,x)}{y(P)},\;
\hat G(P,x)=\f{G_{n+1}(z,x)}{y(P)},\;
\hat H(P,x)=\f{H_n(z,x)}{y(P)}
\lb{479a}
\ea
denote homogeneous quantities encountered in Lemma \ref{l45}.
Since $G(P,x,x')$ is discontinuous at $x=x'$, we introduced the
arithmetic mean of the corresponding one-sided limits following
the usual treatment of first-order systems (see, e.g., \cite{fva},
Sect. 9.4). In fact, the arithmetic mean in (\ref{478a}) leads
to the characteristic function of $D$ (in the terminology of
\cite{fva}, Sect. 9.5), the fundamental object for studying
spectral properties of $D$. The asymptotic expansions (\ref{4.58})
and  (\ref{4.58a}) for $\hat F(P,x)$ and $\hat H(P,x)$ as
$P\to \inf_{\pm}$ then determine the off-diagonal asymptotic
high-energy expansions of of the arithmetic mean of the diagonal
Green's matrix in (\ref{478a}). Similarly, using (\ref{210}) or
(\ref{212}),
\begin{align}
\hat G(P,x)& = \f{1}{2q(x)}[\hat F(P,x)+2iz\hat F_x(P,x)]
 =\f{1}{2p(x)}[\hat H(P,x)-2iz\hat H_x(P,x)] \lb{480a} \\
&   \underset{\zeta \to 0} = \mp 1 + O( \zeta^{-2}) \quad
\text{near} \ \ \infty_{\pm}, \lb{480b}
\end{align}
one obtains the asymptotic high-energy expansion of
(\ref{478a}) as $P\to \inf_{\pm}$ for its diagonal elements.
Even though  (\ref{4.58}), (\ref{4.58a}), (\ref{478a}), and
(\ref{480a}) were derived in the special algebro-geometric
context, we emphasize, however, that the asymptotic expansion of
(\ref{478a}) as $P\to \inf_{\pm}$ only involves the
homogeneous coefficients $\hat f_k(x),\; \hat h_k(x)$
which are universal differential polynomials
in $(p(x),q(x))$.\ Thus, identifying $\Psi_{\pm}(z,x)$ and
$\Psi(P,x,x_0),\; \Psi(P^*,x,x_0)$ as in (\ref{476a})
yields the universal high-energy expansion of the
arithmetic mean of the diagonal Greens matrix $G(z,x,x')$
of $D$ as $z\to \infty$ in the general (not necessarily
algebro-geometric) case. The recursive and hence systematic
approach to this high-energy expansion, based on
(\ref{4.58})--(\ref{459a}), appears to be new.
\end{rem}

The theta function representations for $\phi,\; \Psi,$
and $(p,q)$ then finally read as follows.

\begin{thm}\lb{t46}

Let $P\in\calK_n\backslash\{\infpm\}, \;
(x,x_0,t_r,t_{0,r})\in\bbR^4$,
and assume $\calK_n$ to be nonsingular, that is,
$E_m \neq E_{m'}\text{ for } \;
0 \leq m, m'\leq 2\; n+1.$ Moreover, suppose
$\calD_{\ul {\hat \mu}(x,t)}$, or equivalently,
$\calD_{\ul {\hat \nu}(x,t)}$
to be nonspecial, that is, $i(\calD_{\ul {\hat \mu}(x,t)})=
i(\calD_{\ul {\hat \nu}(x,t)})=0.$ Then
\ba
&&\phi(P,x,t_r)=\f{2i}{q(x_0,t_{0,r})\ome_0}
\f{\tht(\uz_-(\ul {\hat \mu}(x_0,t_{0,r})))}
{\tht(\uz_+(\ul {\hat \mu}(x_0,t_{0,r})))}
\f{\tht(\uz_+(\ul {\hat \mu}(x,t_r)))}
{\tht(\uz_-(\ul {\hat \nu}(x,t_r)))}
\f{\tht(\uz(P,\hat{\ul \nu}(x,t_{r})))}
{\tht(\uz(P,\ul {\hat \mu}(x,t_{r})))}\times \no \\
&&\quad \quad\quad \quad\times
\exp\left[\int_{P_0}^P\ome_{\inf_+, \inf_-}^{(3)}
-2i(x-x_0)e_0-2i(t_r-t_{0,r})\ti e_r \right],
\lb{465}\vspace{2mm} \\
&&\psi_1(P,x,x_0,t_r,t_{0,r})=
\f{\tht(\uz_+(\ul {\hat \mu}(x_0,t_{0,r})))}
{\tht(\uz(P,\ul {\hat \mu}(x_0,,t_{0,r})))}
\f{\tht(\uz(P,\hat{\ul \mu}(x,t_r)))}
{\tht(\uz_+(\ul {\hat \mu}(x,t_r)))} \times \no \\
&&\quad \quad\quad \quad\times
\exp\left[i(x-x_0)\left(e_0+\int_{P_0}^P\Ome_0^{(2)}\right)
+i(t_r-t_{0,r})
\left(\ti e_r+\int_{P_0}^P{\ti \Ome}_r^{(2)}\right)\right],
\lb{466} \vspace{2mm}\\
&&\psi_2(P,x,x_0,t_r,t_{0,r})=
\f{2i}{q(x_0,t_{0,r})\ome_0}
\f{\tht(\uz_-(\ul {\hat \mu}(x_0,t_{0,r})))}
{\tht(\uz(P,\ul {\hat \mu}(x_0,t_{0,r})))}
\f{\tht(\uz(P,\ul {\hat \nu}(x,t_r)))}
{\tht(\uz_-(\ul {\hat \nu}(x,t_r)))}\times \no \\
&&\quad \times
\exp\left[\int_{P_0}^P\ome_{\inf_+, \inf_-}^{(3)}+
i(x-x_0)\left(-e_0+\int_{P_0}^P\Ome_0^{(2)}\right)
+i(t_r-t_{0,r})
\left(-\ti e_r+\int_{P_0}^P{\ti \Ome}_r^{(2)}\right)\right]. \no \\
\lb{467}
\ea
Moreover, one derives
\ba
&&p(x,t_r)=p(x_0,t_{0,r})
\f{\tht(\uz_-(\ul {\hat \nu}(x_0,t_{0,r})))}
{\tht(\uz_+(\ul {\hat \nu}(x_0,t_{0,r})))}
\f{\tht(\uz_+(\ul {\hat \nu}(x,t_r)))}
{\tht(\uz_-(\ul {\hat \nu}(x,t_r)))} \times \no \\
&&\quad \quad\quad \quad\times
\exp[-2i(x-x_0)e_0-2i(t_r-t_{0,r})\ti e_r], \no \\
\lb{468} \\
&&q(x,t_r)=q(x_0,t_{0,r})
\f{\tht(\uz_+(\ul {\hat \mu}(x_0,t_{0,r})))}
{\tht(\uz_-(\ul {\hat \mu}(x_0,t_{0,r})))}
\f{\tht(\uz_-(\ul {\hat \mu}(x,t_r)))}
{\tht(\uz_+(\ul {\hat \mu}(x,t_r)))}\times \no \\
&&\quad \quad\quad \quad\times
\exp[2i(x-x_0)e_0+2i(t_r-t_{0,r})\ti e_r],
\lb{469} \\ \no \\
&&p(x_0,t_{0,r})q(x_0,t_{0,r})=\f{4}{\ome_0^2}
\f{\tht(\uz_+(\ul {\hat \nu}(x_0,t_{0,r})))}
{\tht(\uz_-(\ul {\hat \nu}(x_0,t_{0,r})))}
\f{\tht(\uz_-(\ul {\hat \mu}(x_0,t_{0,r})))}
{\tht(\uz_+(\ul {\hat \mu}(x_0,t_{0,r})))},
\lb{470}
\ea
and
\ba
&& {\ul \al}_{P_0}(\calD_{\ul {\hat \mu}(x,t_r)})=
  {\ul \al}_{P_0}(\calD_{\ul {\hat \mu}(x_0,t_{0,r})})-i(x-x_0)
   \ul U_0^{(2)}-i(t_r-t_{0,r})\ti{\ul U}_r^{(2)},
 \lb{471}\vspace{2mm} \\
&& {\ul \al}_{P_0}(\calD_{\ul {\hat \nu}(x,t_r)})=
  {\ul \al}_{P_0}(\calD_{\ul {\hat \nu}(x_0,t_{0,r})})-i(x-x_0)
   \ul U_0^{(2)}-i(t_r-t_{0,r})\ti{\ul U}_r^{(2)}.
 \lb{472}
\ea
\end{thm}

\begin{proof}
We first prove the $\tht$-function representation (\ref{466}) for
$\psi_1$. Without loss of generality it suffices to treat the
homogeneous case $\hat c_0=1,\; \;\hat c_q=0,\; \;
1\leq q\leq r$. Define the left-hand side of (\ref{466}) to be
 $\ti \psi_1(P,x,x_0,t_r,t_{0,r})$; we need to prove
$\psi_1=\ti \psi_1$ with $\psi_1$ given by (\ref{422}).
For that purpose we first investigate the local zeros
and poles of $\psi_1$. Since they can only come from
zeros of $F_n(z,x_0,s),\; F_n(z,x',t_r)$ in (\ref{422}),
 we note that
\ba
&&q(x',t_r)\phi(P,x',t_r)\underset{P \to \hat \mu_j(x',t_r)}{=}
q(x',t_r)\f{2y(\hat \mu_j(x',t_r))}{-iq(x',t_r)
\prod_{\stackrel{\scriptstyle{k=1}}{\scriptstyle{k \not=j}}}^{n}
(\mu_{j}(x',t_r) - \mu_{k}(x',t_r))}\times
\no \\
&&\quad \quad \times  \f{1}{z-\mu_j(x',t_r)}+O(1)
\underset{P \to \hat \mu_j(x',t_r)}{=}
\f{-\mu_{j,x'}(x',t_r)}{z-\mu_j(x',t_r)}+O(1),
\lb{473}\ea
\ba
&&i\hat F_r(z,x_0,s)\phi(P,x_0,s)
\underset{P \to \hat \mu_j(x_0,s)}{=}
\f{2i\hat F_r(z,x_0,s)y(\hat \mu_j(x_0,s))}{-iq(x_0,s)
\prod_{\stackrel{\scriptstyle{k=1}}{\scriptstyle{k \not=j}}}^{n}
(\mu_j(x_0,s) - \mu_k(x_0,s))}\times
\no \\
&&\quad \quad \times  \f{1}{z-\mu_j(x_0,s)}+O(1)
\underset{P \to \hat \mu_j(x_0,s)}{=}
\f{-\mu_{j,s}(x_0,s)}{z-\mu_j(x_0,s)}+O(1)
\lb{474}
\ea
 using (\ref{415}), (\ref{416}), (\ref{445}), and (\ref{446}).
Thus
\ba
\psi_1(P,x,x_0,t_r,t_{0,r})=
\left\{ \begin{array}{ll}
(z-\mu_j (x,t_r) ) O(1)
&\mbox{ for } P \mbox{ near } \hat
\mu_j (x,t_r) \neq \hat \mu_j (x_0,t_{0,r}),\\ \\
O(1) &\mbox{ for } P \mbox{ near }
 \hat \mu_j (x,t_r) = \hat \mu_j
(x_0,t_{0,r}), \;\\ \\{}
(z-\mu_j(x_0,t_r))^{-1} O(1)
&\mbox{ for } P \mbox{ near }
\hat \mu_j (x_0, t_{0,r}) \neq \hat \mu_j (x, t_r),
\end{array} \right. \no
\ea
\ba
\label{475}
\ea
with $O(1)\neq0$ and hence $\psi_1$ and $\ti \psi_1$ have
identical zeros and poles on $\calK_n\backslash\{\inf_+,
\inf_-\}$ which are all simple. It remains to study the
 behavior of $\psi_1$ near $\inf_{\pm}$. One infers from
(\ref{413})--(\ref{415}) that
\ba
\phi(P,x,t_r)\underset{\zeta \to 0}{=}
\left\{ \begin{array}{ll}
\f{i}{2}p(x,t_r)\zeta+O(\zeta^2),
&P \mbox{ near } \inf_+, \\ \\
\f{2i}{q(x,t_r)}\zeta^{-1}+O(1),
&P \mbox{ near } \inf_-.
\end{array} \right.
\label{476}
\ea
Thus (\ref{415}), (\ref{437}), (\ref{476}), and Lemma 4.5 yield
\ba
&&\int_{x_0}^x dx'[-i\zeta^{-1}+q(x,t_r)\phi(P,x,t_r)]
\no \vspace{2mm}\\
&&+ \int_{t_{0,r}}^{t_r}ds[\ti F_r(\zeta^{-1},x_0,s)
\phi(P,x_0,s)-\ti G_{r+1}(\zeta^{-1},x_0,s)] \no \vspace{2mm}\\
&&\underset{\zeta \to 0}{=}\left[\mp i(x-x_0)\zeta^{-1}+
\left\{ \begin{array}{ll}
O(\zeta),
&P \to \inf_+ \\
O(1),
&P \to \inf_-
\end{array} \right \}\right]\no \vspace{2mm}\\
&& +
\int_{t_{0,r}}^{t_r}ds\left[\f{i\ti F_r(\zeta^{-1},x_0,s)y(P)}
{F_n(\zeta^{-1},x_0,s)}+\f12
\f{F_{n,t_r}(\zeta^{-1},x_0,s)}
{F_n(\zeta^{-1},x_0,s)}\right]\no \vspace{2mm}\\
&& \underset{\zeta \to 0}{=}\left[\mp i(x-x_0)\zeta^{-1}+
\left\{ \begin{array}{ll}
O(\zeta),
&P \to \inf_+ \\
O(1),
&P \to \inf_-
\end{array} \right \}\right]\no \vspace{2mm}\\
&& +
\int_{t_{0,r}}^{t_r}ds\left[\mp i \zeta^{-r-1}
\f{\sum_{\ell=0}^r\ti f_\ell (x_0,s)\zeta^\ell}
{\sum_{\ell=0}^\inf\ti f_\ell (x_0,s)\zeta^\ell}+\f12
\f{q_{t_r}(x_0,s)}
{q(x_0,s)}+O(\zeta)\right]\no \vspace{2mm}\\
&& \underset{\zeta \to 0}{=}\left[\mp i(x-x_0)\zeta^{-1}+
\left\{ \begin{array}{ll}
O(\zeta),
&P \to \inf_+ \\
O(1),
&P \to \inf_-
\end{array} \right \}\right]\no \vspace{2mm}\\
&& +
\int_{t_{0,r}}^{t_r}ds\left[\mp i \zeta^{-r-1}
\pm \f{i\ti f_{r+1} (x_0,s)}
{\ti f_0 (x_0,s)}+\f12
\f{q_{t_r}(x_0,s)}
{q(x_0,s)}+O(\zeta)\right]\no\vspace{2mm}\\
&& \underset{\zeta \to 0}{=}\mp i(x-x_0)\zeta^{-1}
\mp i(t_r-t_{0,r})\zeta^{-r-1}+
\left\{ \begin{array}{ll}
O(\zeta),
&P \to \inf_+, \\ \\
O(1),
&P \to \inf_-,
\end{array} \right.
\lb{477}
\ea
where we used $\ti f_0=-iq$ and
\ba
q_{t_r}=2 \ti f_{r+1}
\lb{478}
\ea
(cf. (\ref{43})) in the homogeneous case
$\ti c_0=1,\; \;\ti c_q=0,\; \;
1\leq q\leq r$. (\ref{477}) yields the correct
essential singularity structure of $\ti \psi_1$
near $\inf_{\pm}$. Moreover, (\ref{342}), (\ref{455}),
and the $O(\zeta)$-term in (\ref{477}) as $P\to \inf_+$
also prove that $\psi_1$ and $\ti \psi_1$ are identically
normalized (near $\inf_+$) and hence coincide by the
$t$-dependent analog of Lemma \ref{l34} (replacing
$-i(x-x_0)\int_{P_0}^P \Ome_0^{(2)}$ by
$-i(x-x_0)\int_{P_0}^P \Ome_0^{(2)}-i(t_r-t_{0,r})
\int_{P_0}^P \ti{\Ome}_r^{(2)})$.
This proves (\ref{466}). The expression (\ref{418}) for
the divisor of $\phi$ then yields
\ba
\phi(P,x,t_r)=C(x,t_r)\f{\tht(\ul z(P,\ul {\hat \nu}(x,t_r)))}
{\tht(\ul z(P,\ul {\hat \mu}(x,t_r)))}e^{\int_{P_0}^{P}
\ome_{\inf_+,\inf_-}^{(3)}},
\lb{479}
\ea
where $C(x,t_r)$ is independent of $P\in\calK_n$.
Thus (\ref{476}) implies
\ba
&& p(x,t_r)=\f{2C(x,t_r)}{i\ome_0}
  \f{\tht(\uz_+(\ul {\hat \nu}(x,t_r)))}
  {\tht(\uz_+(\hat {\ul \mu}(x,t_r)))},
  \label{480} \\
&& q(x,t_r)=\f{2i}{C(x,t_r)\ome_0}
  \f{\tht(\uz_-(\ul {\hat \mu}(x,t_r)))}
  {\tht(\uz_-(\hat {\ul \nu}(x,t_r)))}.
  \label{481}
\ea
Re-examining the asymptotic behavior (\ref{477}) of $\psi_1$
near $\inf_-$, taking into account (\ref{356}), yields
\ba
&&\psi_1(P,x,x_0,t_r,t_{0,r})
\underset{\zeta \to 0}{=}
\f{q(x,t_r)}{q(x_0,t_r)}\exp[i(x-x_0)\zeta^{-1}+O(\zeta)]
\times \no \\
&&\quad \quad \quad \quad \quad \quad \quad \quad \times
\f{q(x_0,t_r)}{q(x_0,t_{0,r})}
\exp[i(t_r-t_{0,r})\zeta^{-1-r}+O(\zeta)] \label{482} \\
&&\underset{\zeta \to 0}{=}
\f{q(x,t_r)}{q(x_0,t_{0,r})}\exp[i(x-x_0)\zeta^{-1}+
i(t_r-t_{0,r})\zeta^{-1-r}+O(\zeta)]\; \;
\mb{ for } P \mb{ near } \inf_-.
\no
\ea
A comparison of (\ref{466}), (\ref{481}), and (\ref{482})
then proves (\ref{469}). A further comparison of (\ref{469})
and (\ref{481})  then determines $C(x,t_r)$ and hence
 yields (\ref{468}) and (\ref{470}). Given $C(x,t_r)$ one
determines $\phi$ in (\ref{465}) from (\ref{479}) and hence
$\psi_2$ in (\ref{467}) from $\psi_2=\phi\psi_1$.
Finally, the linearization property of the Abel map
in (\ref{471}) and (\ref{472}) can be proved directly
using Lagrange interpolation as in the proof of Theorem \ref{t35}.
However, for increasing values of $r$ this method becomes
 exceedingly cumbersome  and it is simpler to resort to
a standard investigation
(cf., e.g., \cite{nmpz}, p. 141--144) of the differential
$\Ome_1(x,x_0,t_r,t_{0,r})=d\ln\psi_1
(.,x,x_0,t_r,t_{0,r})$ (respectively,
$\Ome_2(x,x_0,t_r,t_{0,r})=d\ln\psi_2
(.,x,x_0,t_r,t_{0,r})$ in order to prove (\ref{471})
(respectively, (\ref{472})).
\end{proof}

Since Corollary \ref{c36} extends to the present time-dependent
setting in a straightforward manner we record the
corresponding result without proof,
\begin{equation}
p(x,t_r)q(x,t_r)=-e_{0,1} -
\f{d^2}{dx^2}\ln(\tht(\uz_+(\hat {\ul \mu}(x,t_r)))).
 \lb{483}
\end{equation}

The open constant $q(x_0,t_{0,r})$ in
(\ref{468})--(\ref{470}) is inherent to the \mb{AKNS} formalism
as discussed in Lemma \ref{l36}.

In analogy to Example \ref{e37}, the special case $n=0$
(excluded in Theorem \ref{t46}) yields solutions $(p(x,t_r),
q(x,t_r))$ as in (\ref{468}), (\ref{469}) replacing the theta
quotients by 1.

We note again that the results for $\Psi$ and $(p,q)$ in
Theorem \ref{t46} are known and can be found, for instance, in
\cite{bbei94}, Ch.4, \cite{dub77}, \cite{dub83},
\cite{its1}, \cite{its2}, and \cite{pre85}.
Our main new contribution to this circle of ideas is the
elementary alternative derivation of Theorem \ref{t46} based on the
fundamental meromorphic function $\phi$ on $\calK_n$ and its
connection with the
polynomial recursion formalism of Section \ref{s2}.

\appendix
\section{Hyperelliptic Curves
and Theta Functions}
\lb{app-a}
\renewcommand{\theequation}{\Alph{section}.\arabic{equation}}
\setcounter{prop}{0}
\setcounter{equation}{0}

We briefly summarize our notation and some of the basic facts
on hyperelliptic curves and their theta functions as employed
in Sections 3 and 4. For background information
on this standard material we refer, for instance, to
\cite{fk}, Chs. I--III, IV, \cite{fay}, \cite{gh}, Ch. 2,
\cite{kra}, Ch. X, \cite{mum1}, Ch. 2.

Consider the points
\ba
\{E_m\}_{0\leq m \leq 2n+1} \subset \bbC,\; \;
n\in\bbN_0
\lb{a1}
\ea
and introduce an appropriate set of $n+1$ (nonintersecting)
cuts $\calC_j$ joining $E_j$ and $E_{k(j)}$, where
$E_{k(j)}=E_j$ for some $j$ is permitted in order to include
singular curves. Denote
\ba
\calC=\underset{j \in J}{\bigcup} \calC_j, \; \;
\calC_j\cap \calC_k=\emptyset \text{ for } j\neq k,
\lb{a2}
\ea
where the finite index set $J \subset \{0,1,\ldots,2n+1\}$
has cardinality $n+1$ and define the cut plane $\Pi$,
\ba
\Pi:=\bbC \; \backslash \; \calC.
\lb{a3}
\ea
Next, introduce the holomorphic function
\ba
{R}_{2n+2} (.)^{1/2} : \left\{\bma{l}
\Pi \to \bbC \\
z\mapsto [\prod_{m=0}^{2n+1} (z-E_m)]^{1/2}
\lb{a4}
\ema \right.
\ea
on $\Pi$ with a definite choice of the square root branch
(\ref{a4}). Given the holomorphic function (\ref{a4}) one
defines the set
\begin{equation}
M_n=\{(z,\sig R_{2n+2} (z)^{1/2})\;|\; z\in\bbC,
\, \sig\in \{+,-\}\} \cup \{
\infty_+, \infty_-\}
\lb{a5}
\end{equation}
and
\ba
B_s=\{(E_m,0)\}_{0\leq m \leq 2n+1},
\lb{a6}
\ea
the set of branch and/or singular points. $M_n$ becomes a
Riemann surface upon introducing appropriate charts
 $(U_{P_0},\zeta_{P_0})$
 defined in a standard manner. Let
\begin{align}
\begin{split}
P_0 &  = (z_0, \sig_0 R_{2n+2} (z_0)^{1/2})
\text{ or } P_0 =\infty_\pm, \\ \lb{a7} \\
P&=(z,\sig R_{2n+2} (z)^{1/2})
 \in U_{P_0} \subset M_n,\;
V_{P_0}  = \zeta_{P_0} (U_{P_0})
 \subset \bbC.
\end{split}
\end{align}
\underline{$P_0  \notin \{B_s\cup \{\infty_+,
\infty_-\}\}$:}
\begin{align}
U_{P_0} = & \{P\in M_n \; \big|\; |z-z_0| < C_{z_0},
\; \sig R_{2n+2} (z)^{1/2} \text{
the branch obtained by straight line} \no \\ & \text{analytic
continuation starting from } z_0\}, \quad C_{z_0} =
\underset{m}{\min} | z_0 - E_m|, \no \\
V_{P_0} = & \{\zeta\in\bbC\; \big| \; |\zeta| < C_{z_0}\}, \no
\end{align}
\begin{align}
\begin{split}
& \zeta_{P_0} : \left\{
\begin{array}{l} U_{P_0} \to V_{P_0}\\
P \mapsto (z-z_0),
\end{array} \right.
\quad \zeta_{P_0}^{-1}: \left\{ \begin{array}{l}
V_{P_0} \to U_{P_0}\\
\zeta \mapsto (z_0 +\zeta, \, \sig R_{2n+2} (z_0
+\zeta)^{1/2}). \end{array} \right.
\lb{a8}
\end{split}
\end{align}
\underline{$P_0 = \infty_\pm$:}
\begin{align}
\begin{split}
& U_{P_0} = \{P\in M_n\; \big|\; |z|> C_{\infty}\},\: C_{\infty}
=\underset{m}{\max} |E_m|, \:
V_{P_0} =\big\{ \zeta\in\bbC\; \big|\;
| \zeta| < C^{-1}_{\infty}\big\},  \\
&\zeta_{P_0}: \left\{ \begin{array}{l}
U_{P_0} \to V_{P_0}\\
P\mapsto z^{-1}\\
\infty_\pm \mapsto 0,
\end{array}\right. \quad
\zeta_{P_0}^{-1} : \left\{
\begin{array}{l}
V_{P_0} \to U_{P_0}\\
\zeta \mapsto (\zeta^{-1} ,
\mp  [\Pi_m (1-
E_m\zeta)]^{1/2}\zeta^{-n-1})\\
0 \mapsto \infty_\pm,
\end{array}\right.
\\
&{\scriptstyle [ \Pi_{m} (1-E_m\zeta )]^{1/2}
=1 -{{\f12} ({ \sum_{m}} E_m)}\zeta
+O
(\zeta^2)}.
\lb{a9}
\end{split}
\end{align}

Similarly, local coordinates for branch and/or singular
points $P_0 \in B_s$ are defined as
$\zeta_{P_0}(P)=(z-z_0)^{r/2}$ for appropriate
$r=1$ or $2$. For the reader's
convenience we provide a detailed treatment of branch
points in the nonsingular case (where $E_m \neq E_{m'}
\text{ for } m \neq m'$) for the two most frequently occurring
situations, the self-adjoint case where $\{E_m\}_{0\leq m
\leq 2n+1} \subset \bbR$ and the case where $\{E_m\}_{0\leq m
\leq 2n+1}=\{\epsilon_{\ell},\ol{\epsilon_{\ell}}\}_{0\leq \ell
\leq n}$ consists of complex conjugate pairs at the end of
this appendix.

In addition, it is useful to consider the subsets $\Pi_{\pm}
\subset M_n$ (i.e., upper and lower sheets)
\begin{equation}
\Pi_\pm = \{ (z, \pm R_{2n+2} (z)^{1/2})
\in M_n\; |\; z\in\Pi\}
\lb{a10}
\end{equation}
and the associated charts
\begin{equation}
\zeta_\pm : \left\{ \begin{array}{l}
\Pi_\pm \to \Pi\\
P \mapsto z
\end{array} \right..
\lb{a11}
\end{equation}
(\ref{a8}), (\ref{a9}), and the corresponding charts for
$P_0\in B_s$ define a complex structure on $M_n$. We shall
denote the resulting Riemann surface by $\calK_n$. In general,
$\calK_n$ is a (possibly singular) curve of (arithmetic)
genus $n$.

Next, consider the holomorphic sheet exchange
map (involution)
\begin{equation}
* : \left\{ \begin{array}{l}
\calK_n \to \calK_n\\
(z,\sig R_{2n+2} (z)^{1/2})
\mapsto (z, \sig R_{2n+2} (z)^{1/2})^* =(z,-\sig
R_{2n+2} (z)^{1/2})\\
\infty_\pm \mapsto \infty_\pm^* =
 \infty_\mp \end{array} \right.
\lb{a12}
\end{equation}
and the two meromorphic projection maps
\begin{equation}
\ti \pi : \begin{cases}
\calK_n \to \bbC \cup \{\infty\}\\
(z, \sig R_{2n+2} (z)^{1/2}) \mapsto z\\
\infty_\pm \mapsto \infty,
\end{cases} \quad
R_{2n+2}^{1/2} : \begin{cases}
\calK_n \to \bbC\cup \{\infty\}\\
(z, \sig R_{2n+2} (z)^{1/2}) \mapsto
 \sig R_{2n+2}(z)^{1/2}\\
\infty_\pm \mapsto \infty.
\end{cases}
\lb{a13}
\end{equation}
$\ti \pi$ has poles of order 1 at $\infty_\pm$
and $R_{2n+2} (z)^{1/2}$ has poles of order
$n+1$ at $\infty_\pm$.  Moreover,
\begin{equation}
\ti \pi (P^*) =\ti \pi (P), \; R_{2n+2}^{1/2}
(P^*) =-R^{1/2}_{2n+2} (P),
\quad P \in \calK_n.
\lb{a14}
\end{equation}
Thus $\calK_n$ is a two-sheeted ramified covering of
the Riemann sphere
$\bbC_\infty (\cong \bbC \cup \{ \infty\}$),
$\calK_n$ is compact (since
$\ti\pi$ is open and $\bbC_\infty$ is compact), and $\calK_n$ is
hyperelliptic (since it admits the meromorphic
function $\ti\pi$ of
degree two).

In the following we abbreviate
\ba
P=(z,y),\; \;P\in \calK_n \backslash \{\infpm\},
\lb{a15}
\ea
(i.e., we define $y(P)=R^{1/2}_{2n+2} (P),$ see (\ref{a13})).

Next we turn to nonsingular curves $\calK_n$ where
\ba
E_m\neq E_{m'},\text{ for }m\neq m',\; 0\leq m,m'\leq 2n+1.
\lb{a16}
\ea
One infers that for
$n\in \bbN$, $d\ti\pi /
y$ is a holomorphic differential
on $\calK_n$ with zeros of
order $n-1$ at $\infty_\pm$ and hence
\begin{equation}
\eta_j =\dfrac{\ti\pi^{j-1} d\ti\pi}{y},
\quad 1\leq j \leq n
\lb{a17}
\end{equation}
form a basis for the space of holomorphic
differentials on $\calK_n$.

Next we introduce a canonical homology basis
$\{a_j, b_j\}_{1\leq j \leq
n}$ for $\calK_n$ where the cycles
 are chosen such that their
intersection matrix reads
\begin{equation}
a_j \circ b_k =\del_{j,k}, \quad 1\leq j,k \leq n.
\lb{a18}
\end{equation}
Introducing the invertible matrix $C$ in $\bbC^n$,
\begin{align}
\begin{split}
& C=(C_{j,k})_{1\leq j,k\leq n}, \; C_{j,k}
=\int_{a_k} \eta_j,
\\ \lb{a19} \\
& \ul c (k) =(c_1(k), \ldots, c_n(k)), \; c_j (k)
=(C^{-1})_{j,k},
\end{split}
\end{align}
the normalized differentials $\ome_j$,
$1\leq j \leq n$,
\begin{equation}
\ome_j =\sum_{\ell=1}^n c_j (\ell) \eta_\ell,
\; \int_{a_k} \ome_j
=\del_{j,k}, \quad 1\leq j,k\leq n
\lb{a20}
\end{equation}
form a canonical basis for the space of
holomorphic differentials on
$\calK_n$.  The matrix $\tau$ in $\bbC^n$ of
$b$-periods,
\begin{equation}
\tau=(\tau_{j,k})_{1\leq j,k\leq n}, \quad \tau_{j,k}
 =\int_{b_k}\ome_j
\lb{a21}
\end{equation}
satisfies
\begin{equation}
\tau_{j,k} =\tau_{k,j}, \quad 1\leq j,k\leq n,
\lb{a22}
\end{equation}
\begin{equation}
\text{Im}(\tau) =\f{1}{2i}(\tau-\tau^{*})>0.
\lb{a23}
\end{equation}

In the charts $(U_{\infty_\pm},
\zeta_{\infty_\pm} \equiv \zeta)$
induced by $1/ \ti\pi$ near $\infty_\pm$ one
infers
\begin{align}
\begin{split}
\ul \ome & = \pm \sum_{j=1}^n \ul c (j) \f{\zeta^{n-j}\,
d\zeta}{[\Pi_m
(1- E_m\zeta)]^{1/2}}\\
& = \pm \Big\{ \ul c (n)
+\Big[ \frac12 \ul c (n) \sum_{m=0}^{2n+1} E_m
+\ul c (n-1)\Big]\zeta  +O(\zeta^2) \Big\} \, d\zeta.
\lb{a24}
\end{split}
\end{align}

Associated with the homology basis
$\{a_j, b_j\}_{1\leq j \leq n}$ we
also recall the canonical dissection of $\calK_n$
along its cycles yielding
the simply connected interior $\hat \calK_n$ of the
fundamental polygon $\pa
\hat \calK_n$ given by
\begin{equation}
\partial  \hat \calK_n =a_1 b_1 a_1^{-1} b_1^{-1}
a_2 b_2 a_2^{-1} b_2^{-1} \cdots
a_n^{-1} b_n^{-1}.
\lb{a25}
\end{equation}

The Riemann theta function associated with
$\calK_n$ is defined by
\begin{equation}
\theta (\ul z) =\sum_{\ul n \in\bbZ^n}
 \exp [2\pi i (\ul n,\ul z) + \pi i (\ul n,
\tau \ul n)], \quad \ul z =(z_1,
\ldots, z_n) \in\bbC^n,
\lb{a26}
\end{equation}
where $(\ul u, \ul v)=\sum_{j=1}^n \overline{u}_j v_j$
denotes the
scalar product
in $\bbC^n$. It has the fundamental properties
\begin{align}
\begin{split}
& \theta(z_1, \ldots, z_{j-1}, -z_j, z_{j+1},
\ldots, z_n) =\theta
(\ul z),\\
& \theta (\ul z +\ul m +\tau \ul n)
=\exp [-2 \pi i (\ul n,\ul z) -\pi i (\ul n, \tau
\ul n) ] \theta (\ul z), \quad \ul m, \ul n \in\bbZ^n.
\lb{a27}
\end{split}
\end{align}

A divisor $\calD$ on $\calK_n$ is a map
$\calD: \calK_n \to \bbZ$, where $\calD
(P) \neq 0$ for only finitely many
$P\in \calK_n$.  The set of all divisors
on $\calK_n$ will be denoted by $\Div(\calK_n)$.
With $L_n$ we denote the
period lattice
\begin{equation}
L_n : = \{ \ul z \in\bbC^n \; |\;  \ul z = \ul m +\tau \ul n,
\; \ul m, \ul n \in\bbZ^n\}
\lb{a28}
\end{equation}
and the Jacobi variety $J(\calK_n)$ is defined by
\begin{equation}
J(\calK_n) =\bbC^n / L_n.
\lb{a29}
\end{equation}
The Abel maps $\ul A_{P_0} (.)$ respectively
$\ul \al_{P_0} (.)$ are defined by
\begin{align}
\ul A_{P_0} & : \begin{cases}
\calK_n \to J(\calK_n)\\
P\mapsto \ul A_{P_0} (P) =\int_{P_{0}}^P \ul \ome \mod (L_n),
\end{cases}
\lb{a30}\\
\ul \al_{P_0} & : \begin{cases}
\Div(\calK_n) \to J(\calK_n)\\
\calD \mapsto \ul \al_{P_0} (\calD)
=\sum_{P \in \calK_n} \calD (P) \ul A_{P_0} (P),
\end{cases}
\lb{a31}
\end{align}
with $P_0 \in \calK_n$ a fixed base point.
(In the main text we agree
to fix ${P_0} =(E_0, 0)$ for convenience.)

In connection with (\ref{a25}) we shall also
need the maps (cf. (\ref{334}))
\begin{equation}
\ul{\hat A}_{P_0} : \begin{cases}
\hat \calK_n \to \bbC^n\\
P\mapsto \int_{P_0}^P \ul \ome,
\end{cases}
\quad \hat{\ul \al}_{P_0}\ : \begin{cases}
\Div(\calK_n) \to \bbC^n\\
\calD \mapsto \sum_{P\in\hat \calK_n} \calD (P)
 \ul{\hat A}_{P_0} (P),
\end{cases}
\lb{a32}
\end{equation}
with path of integration lying in $\hat \calK_n$.

Let $\calM (\calK_n)$ and $\calM^1 (\calK_n)$ denote the
set of meromorphic
functions (0-forms) and meromorphic
differentials (1-forms)
on $\calK_n$. The residue of a meromorphic differential
$\nu\in \calM^1 (\calK_n)$ at a
point $Q_0 \in \calK_n$ is defined by
\begin{equation}
\res_{Q_0} (\nu)
=\frac{1}{2\pi i} \int_{\gam_{Q_0}} \nu,
\lb{a33}
\end{equation}
where $\gam_{Q_0}$ is a counterclockwise oriented
smooth simple closed
contour encircling $Q_0$ but no other pole of
$\nu$.  Holomorphic
differentials are also called Abelian differentials
of the first kind (dfk). Abelian differentials of the second kind
(dsk) $\ome^{(2)} \in \calM^1 (\calK_n)$ are characterized
by the property
that all their residues vanish.  They are
normalized, for
instance, by demanding that all their $a$-periods
vanish, that is,
\begin{equation}
\int_{a_j} \ome^{(2)} =0, \quad 1\leq j \leq n.
\lb{a34}
\end{equation}
If $\ome_{P_1, n}^{(2)}$ is a dsk on $\calK_n$ whose
only pole is $P_1 \in \hat \calK_n$ with principal part
$\zeta^{-n-2}\,d\zeta$, $n\in\bbN_0$ near
$P_1$ and $\ome_j =
 (\sum_{m=0}^\infty d_{j,m} (P_1) \zeta^m)\, d\zeta$
near $P_1$, then
\begin{equation}
\int_{b_j} \ome_{P_1, n}^{(2)} =
 \frac{2\pi i}{n+1} d_{j,n} (P_1).
\lb{a35}
\end{equation}

Any meromorphic differential $\ome^{(3)}$ on
$\calK_n$ not of the first or
second kind is said to be of the third
kind (dtk).
A dtk $\ome^{(3)} \in \calM^1 (\calK_n)$
is usually normalized by the vanishing of its
$a$-periods, that is,
\begin{equation}
\int_{a_j} \ome^{(3)} =0, \quad 1\leq j\leq n.
\lb{a36}
\end{equation}
A normal dtk $\ome_{P_1, P_2}^{(3)}$ associated
with two points $P_1$,
$P_2 \in \hat \calK_n$, $P_1 \neq P_2$ by definition
has simple poles at
$P_1$ and $P_2$ with residues $+1$ at $P_1$ and
$-1$ at $P_2$ and
vanishing $a$-periods.  If $\ome_{P,Q}^{(3)}$ is a
normal dtk associated
with $P$, $Q\in\hat \calK_n$, holomorphic on
$\calK_n \backslash \{ P,Q\}$, then
\begin{equation}
\int_{b_j} \ome_{P,Q}^{(3)} =2\pi i \int_{Q}^P \ome_j,
\quad 1\leq j \leq n,
\lb{a37}
\end{equation}
where the path from $Q$ to $P$ lies in
$\hat \calK_n$ (i.e.,
does not touch any of the cycles $a_j$, $b_j$).

We shall always assume (without loss of generality)
that all poles of
dsk's and dtk's on $\calK_n$ lie on $\hat \calK_n$ (i.e.,
not on $\pa \hat \calK_n$).

For $f\in \calM (\calK_n) \backslash \{0\}$,
$\ome \in \calM^1 (\calK_n) \backslash \{0\}$ the
divisors of $f$ and $\ome$ are denoted
by $(f)$ and
$(\ome)$, respectively.  Two
divisors $\calD$, $\calE\in \Div(\calK_n)$ are
called equivalent, denoted by
$\calD \sim \calE$, if and only if $\calD -\calE
=(f)$ for some
$f\in\calM (\calK_n) \backslash \{0\}$.  The divisor class
$[\calD]$ of $\calD$ is
then given by $[\calD]
=\{\calE \in \Div(\calK_n)\; | \; \calE \sim \calD\}$.  We
recall that
\begin{equation}
\deg ((f))=0,\, \deg ((\ome)) =2(n-1),\,
f\in\calM (\calK_n) \backslash
\{0\},\,  \ome\in \calM^1 (\calK_n) \backslash \{0\},
\lb{a38}
\end{equation}
where the degree $\deg (\calD)$ of $\calD$ is given
by $\deg (\calD)
=\sum_{P\in \calK_n} \calD (P)$.  It is custom to call
$(f)$ (respectively,
$(\ome)$) a principal (respectively, canonical)
divisor.

Introducing the complex linear spaces
\begin{align}
\calL (\calD) & =\{f\in \calM (\calK_n)\; |\; f=0
 \text{ or } (f) \geq \calD\}, \;
r(\calD) =\dim_\bbC \calL (\calD),
\lb{a39}\\
\calL^1 (\calD) & =
 \{ \ome\in \calM^1 (\calK_n)\; |\; \ome=0
 \text{ or } (\ome) \geq
\calD\},\; i(\calD) =\dim_\bbC \calL^1 (\calD),
\lb{a40}
\end{align}
($i(\calD)$ the index of specialty of $\calD$) one
infers that $\deg
(\calD)$, $r(\calD)$, and $i(\calD)$ only depend on
the divisor class
$[\calD]$ of $\calD$.  Moreover, we recall the
following fundamental
facts.

\begin{thm} \lb{ta1}
Let $\calD \in \Div(\calK_n)$,
$\ome \in \calM^1 (\calK_n) \backslash \{0\}$. Then\\
(i).
\begin{equation}
i(\calD) =r(\calD-(\ome)), \quad n\in\bbN_0.
\lb{a41}
\end{equation}
(ii) (Riemann-Roch theorem).
\begin{equation}
r(-\calD) =\deg (\calD) + i (\calD) -n+1,
\quad n\in\bbN_0.
\lb{a42}
\end{equation}
(iii) (Abel's theorem).  $\calD\in \Div(\calK_n)$,
$n\in\bbN$ is principal
if and only if
\begin{equation}
\deg (\calD) =0 \text{ and } \ul \al_{P_0} (\calD)
=\ul{0}.
\lb{a43}
\end{equation}
(iv) (Jacobi's inversion theorem).  Assume
$n\in\bbN$, then $\ul \al_{P_0}
: \Div(\calK_n) \to J(\calK_n)$ is surjective.
\end{thm}

For notational convenience we agree to abbreviate
\begin{equation}
\calD_Q: \begin{cases}
\calK_n \to \{0,1\}\\
P \mapsto \begin{cases}
1, & P=Q\\
0, & P\neq Q
\end{cases}
\end{cases}
\lb{a44}
\end{equation}
and, for $\ul Q =(Q_1, \ldots, Q_n) \in \sig^n \calK_n$
($\sig^n \calK_n$ the $n$-th symmetric power of $\calK_n$),
\begin{equation}
\calD_{\ul Q} : \begin{cases}
\calK_n \to \{0,1,\ldots, n\}\\
P \mapsto { \begin{cases}
k & \text{if }
P \text{ occurs } k \text{ times in } \{Q_1, \ldots,
Q_n\}\\
0 & \text{if } P \notin \{Q_1,\ldots, Q_n\}.
\end{cases}}
\end{cases}
\lb{a45}
\end{equation}
Moreover, $\sig^n \calK_n$ can be identified with
the set of positive
divisors $0< \calD \in \Div(\calK_n)$ of degree $n$.

\begin{lem} \lb{la2}
Let $\calD_{\ul Q} \in \sig^n \calK_n$,
$\ul Q=(Q_1, \ldots, Q_n)$.  Then
\begin{equation}
1 \leq i (\calD_{\ul Q} ) =s(\leq n/2)
\lb{a46}
\end{equation}
if and only if there are $s$ pairs of the type
$(P, P^*)\in \{Q_1,
\ldots, Q_n\}$ (this includes, of course, branch
points for which
$P=P^*$).
\end{lem}

Finally, still assuming the nonsingular case (\ref{a16})
for simplicity, we consider two frequently encountered special
cases, namely

\ul{Case I}: The self-adjoint case, where
\ba
\{E_m\}_{0\leq m\leq 2n+1} \subset \bbR,\; \;
E_0<E_1< \ldots <E_{2n+1}
\lb{a47}
\ea
and

\ul{Case II}:  Complex conjugate pairs of branch points, that is,
\ba
\{E_m\}_{0\leq m\leq 2n+1}=
\{\epsilon_{\ell},\ol{\epsilon_{\ell}}\}_{0\leq \ell
\leq n}.
\lb{a48}
\ea
Without loss of generality we assume
\ba
&&\text{Re}(\epsilon_{\ell}) < \text{Re}(\epsilon_{\ell +1}),\;
0\leq \ell \leq n-1,\;
\text{Im}(\epsilon_{\ell})< \text{Im}(\ol{\epsilon_{\ell}}),\;
0\leq \ell \leq n.
\lb{a48a}
\ea
We start with

\ul{Case I}: Define
\ba
\calC_j=[E_{2j},E_{2j+1}],\; \;
0\leq j \leq n,
\lb{a49}
\ea
and extend ${R}_{2n+2}(.)^{1/2}$ in (\ref{a4}) to all
of $\bbC$ by
\begin{equation}
{R}_{2n+2} (\lam)^{1/2}
=\lim\limits_{\eps \downarrow 0} {R}_{2n+2} (\lam +
i\eps)^{1/2}, \quad \lam \in \calC,
\lb{a50}
\end{equation}
with the sign of the square root chosen
according to
\begin{equation}
 {R}_{2n+2} (\lam)^{1/2} =
| {R}_{2n+2} (\lam)^{1/2}|
\left\{ \begin{array}{ll}
-1,
& \lam \in (E_{2n+1}, \infty), \\
(-1)^{n+j+1},
& \lam \in (E_{2j+1}, E_{2j+2}),
\; 0 \leq j \leq n-1,  \\
(-1)^n,
& \lam \in (-\infty, E_0), \lb{a50a} \\
i(-1)^{n+j+1}  ,
& \lam \in (E_{2j}, E_{2j+1}), \;
0\leq j \leq n.
\end{array} \right.
\end{equation}

In this case (\ref{a8}) and (\ref{a9}) are supplemented as follows.

\underline{$P_0 = (E_{m_0}, 0)$:}
\begin{align}
& U_{P_0} = \big\{P\in M_n\; \big|\;  | z-E_{m_0}|
< C_{m_0}\big\}, \; C_{m_0}
=\underset{m\neq m_0}{\min} | E_{m_0} - E_m|,\no \\
& V_{P_0} =\{\zeta\in\bbC \;
\big|\;  | \zeta| < C^{1/2}_{m_0} \}, \no
\end{align}
\begin{align}
\begin{split}
& \zeta_{P_0} : \left\{ \begin{array}{l}
U_{P_0} \to V_{P_0}\\
P\mapsto \sig (z-E_{m_0})^{1/2},\end{array} \right. \quad
\begin{array}{l} (z-E_{m_0})^{1/2}
=|(z-E_{m_0})^{1/2}|
 e^{(i/2)\arg (z-E_{m_0})},\\
\arg(z-E_{m_0}) \in \begin{cases} [0, 2\pi),
& m_0 \text{ even},\\
(-\pi, \pi], & m_0 \text{ odd },
\end{cases} \end{array} \\
& \zeta_{P_0}^{-1} : \left\{ \begin{array}{l}
V_{P_0} \to U_{P_0}\\
\zeta \mapsto (E_{m_0} +\zeta^2, \;
[\prod_{m\neq m_0} (E_{m_0} -E_m
+\zeta^2)]^{1/2}\zeta),\end{array} \right.\\
& \scriptstyle{\left[\prod_{m\neq m_0} (E_{m_0} -E_m
+\zeta^2)\right]^{1/2} =(-1)^n
i^{-m_0-1}\;
\big| \left[ \prod_{m\neq m_0}
 (E_{m_0} -E_m)\right]^{1/2} \big|}
\scriptstyle{\left[1
+\frac12  \left(\sum_{m\neq m_0} (E_{m_0}
-E_m)^{-1}\right)\zeta^2 +O(\zeta^4)\right]}.\lb{a51}
\end{split}
\end{align}

\ul{Case II}:\ Define
\ba
\calC_\ell=\{z\in \bbC\; |\;
z=\epsilon_{\ell}+t(\ol{\epsilon_{\ell}}-
\epsilon_{\ell}),\; 0\leq t \leq 1\},
\; \; 0\leq \ell \leq n
\lb{a52}
\ea
and extend ${R}_{2n+2}(.)^{1/2}$ in (\ref{a4}) to all
of $\bbC$ by
\begin{equation}
{R}_{2n+2} (z)^{1/2}
=\lim\limits_{\eps \downarrow 0} {R}_{2n+2} (z +(-1)^{n+\ell}
\eps)^{1/2}, \quad z \in \calC_{\ell},\; 0\leq \ell \leq n,
\lb{a53}
\end{equation}
with the sign of the square root chosen
according to
\begin{equation}
 {R}_{2n+2} (\lam)^{1/2} =
| {R}_{2n+2} (\lam)^{1/2}|
\left\{ \begin{array}{ll}
-1,
& \text{Re}(\lam) \in (\epsilon_{n}, \infty), \\
(-1)^{n+\ell+1},
& \lam \in (\text{Re}(\epsilon_{\ell}),
\text{Re}(\epsilon_{\ell+1})),
\; 0 \leq \ell \leq n-1,  \\
(-1)^n,
& \lam \in (-\infty, (\text{Re}(\epsilon_0)).
\end{array} \right. \lb{a54}
\end{equation}
In this case (\ref{a8}) and (\ref{a9}) are supplemented
as follows.

\underline{$P_0 = (E_{m_0}, 0)$}:
\begin{align}
& U_{P_0} = \big\{P\in M_n\; \big|\;  | z-E_{m_0}|
< C_{m_0}\big\}, \; C_{m_0}
=\underset{m\neq m_0}{\min} | E_{m_0} - E_m|,\no \\
& V_{P_0} =\{\zeta\in\bbC \;
\big|\;  | \zeta| < C^{1/2}_{m_0} \}, \no
\end{align}
\ba
\zeta_{P_0} : \left\{ \begin{array}{ll}
U_{P_0} \to V_{P_0}\\
P\mapsto \sig (z-E_{m_0})^{1/2},\end{array} \right.\quad
(z-E_{m_0})^{1/2}
=|(z-E_{m_0})^{1/2}|
 e^{(i/2)\arg (z-E_{m_0})},
\lb{a55} \quad
\ea

\begin{align}
&\arg (z-\epsilon_{\ell})\in
\left\{ \begin{array}{l}
(\f{\pi}{2},\f{5\pi}{2}],\;\ell\; \text{even}, \\  \\{}
[\f{\pi}{2},\f{5\pi}{2}),\;\ell\; \text{odd},
\end{array}\right.\quad
\arg (z-\ol{\epsilon_{\ell}})\in
\left\{ \begin{array}{l}
[-\f{\pi}{2},\f{3\pi}{2}),\;\ell\; \text{even}, \\  \\
(-\f{\pi}{2},\f{3\pi}{2}],\;\ell\; \text{odd},
\end{array}\right.
\quad n \; \text{even},\no  \\ \no \\
&\arg (z-\epsilon_{\ell})\in
\left\{ \begin{array}{l}
[ \f{\pi}{2},\f{5\pi}{2} ),\;\ell\; \text{even}, \\  \\{}
(\f{\pi}{2},\f{5\pi}{2} ],\;\ell\; \text{odd},
\end{array}\right. \quad
\arg (z-\ol{\epsilon_{\ell}})\in
\left\{ \begin{array}{l}
(-\f{\pi}{2},\f{3\pi}{2}],\;\ell\; \text{even}, \\ \\{}
[-\f{\pi}{2},\f{3\pi}{2}),\;\ell\; \text{odd},
\end{array}\right.
\quad n \; \text{odd},\no
\end{align}

\ba
&& \zeta_{P_0}^{-1} : \left\{ \begin{array}{ll}
V_{P_0} \to U_{P_0}\\
\zeta \mapsto (E_{m_0} +\zeta^2, \;
[\prod_{m\neq m_0} (E_{m_0} -E_m
+\zeta^2)]^{1/2}\zeta), \end{array} \right.
\quad\quad\quad\quad\quad\quad\quad\quad\quad\quad\quad\quad\no \\
&&{\scriptstyle{ \left[\prod_{m\neq m_0} (E_{m_0} -E_m
+\zeta^2)\right]^{1/2} =e^{(i/2)\sum_{m\neq m_0} \arg(E_{m_0}-E_m)}
\;
\big| \left[ \prod_{m\neq m_0}
 (E_{m_0} -E_m)\right]^{1/2}\big|\times}} \no\\
&&\quad\quad\quad\quad\quad\quad\quad\quad\quad\quad\quad\quad
\quad\quad\quad\quad\quad\quad{\scriptstyle{  \times
\left[1
+\frac12  \left(\sum_{m\neq m_0} (E_{m_0}
-E_m)^{-1}\right)\zeta^2 +O(\zeta^4)\right]}},\no
\ea
where $\exp[(i/2)\sum_{m\neq m_0} \arg(E_{m_0}-E_m)]$ can
be determined from (\ref{a54}) by analytic continuation.

Cases I and II are of course compatible with our general choice of
\ba
y(P)=R_{2n+2}^{1/2}(P)\underset{\zeta \to 0}{=}
\mp\Big[ 1
-\frac12  \Big(\sum_{m=0}^{2n+1} E_{m}\Big)\zeta
 +O(\zeta^2)\Big] \zeta^{-n-1} \text{ as } P\to \inf_{\pm}.
\lb{a56}
\ea


\section{An Explicit Illustration of the Riemann-Roch Theorem}
\lb{app-b}
\setcounter{prop}{0}
\setcounter{equation}{0}

We provide a brief illustration of the Riemann-Roch theorem in
connection with nonsingular hyperelliptic curves $\calK_n$ of
the type
(\ref{222}) and explicitly determine a basis for the vector space
$\calL (-k\calD_{\infty_{-}} - m(k)\calD_{\infty_{+}}
- \calD_{\hat{\ul \mu} (x_0)} )$, where
$m(k)=\text{max}\,(0,k-2)$ and $k\in\bbN_0$.
(The corresponding case of hyperelliptic curves $\calK_n$ branched
at infinity has been discussed in Appendix B of \cite{grt}.)

We freely use the notation introduced in
Appendix~A and refer, in
particular, to the definition (\ref{a39})
of $\calL (\calD)$ and the
Riemann-Roch theorem stated in
Theorem~\ref{ta1} (ii).  In addition, we
use the short-hand notation
\begin{align}
k \calD_{\infty_{-}} + m(k) \calD_{\infty_{+}}
+ \calD_{\hat{\ul \mu} (x_0)} =
& \sum_{\ell=1}^k \calD_{\infty_{-}}
+ \sum_{\ell=1}^{m(k)} \calD_{\infty_{+}}
+ \sum_{j=1}^n \calD_{\hat \mu_j (x_0)}, \lb{b1} \\
&k\in\bbN_0,\; \;\ul{\hat \mu}(x_0)=({\hat \mu}_1(x_0), \ldots,
{\hat \mu}_n(x_0)) \notag
\end{align}
and recall that
\begin{align}
&\calL (-k\calD_{\infty_{-}} - m(k)\calD_{\infty_{+}}
-\calD_{\hat{\ul \mu}(x_0)}) \notag \\
&= \{ f\in\calM (\calK_n)\; \big|\;  f=0
\mbox{ or } (f) + k\calD_{\infty_{-}} +
m(k)\calD_{\infty_{+}} +
 \calD_{\hat{\ul \mu} (x_0)} \geq 0\}, \quad
k \in\bbN_0.
\lb{b2}
\end{align}
With $\phi (P, x)$, $\psi_j (P,x,x_0),\; j=1,2$
defined as in (\ref{38}), (\ref{310}),
(\ref{323}) we obtain the following result.

\begin{thm} \lb{tb1}
Assume $\calD_{\hat{\ul \mu}(x_0)}$ to be nonspecial
(i.e., $i( \calD_{\hat{\ul \mu}
(x_0) } ) =0$) and of degree $n\in \bbN$.
For $k\in\bbN_0$, a basis for
the vector space $\calL (-k \calD_{\infty_{-}}
-m(k)\calD_{\infty_{+}} - \calD_{\hat{\ul \mu} (x_0)})$
is given by
\begin{equation}
\begin{cases}
\{1\}, & k=0, \\
\{ \ti \pi^\ell\}_{0 \leq \ell \leq m(k)}
\cup \{ \ti\pi^\ell \phi (.,x_0)
\}_{0\leq \ell \leq  k-1}, & k \in \bbN.
\end{cases}
\lb{b3}
\end{equation}
\end{thm}

\begin{proof}
The elements in (\ref{b3}) are easily seen to
be linearly independent and
belonging to $\calL (-k\calD_{\infty_{-}}
-m(k)\calD_{\infty_{+}}
 - \calD_{\hat{\ul \mu} (x_0)} )$.  It
remains to be shown that they are maximal.
Since $i( \calD_{\hat{\ul \mu} (x_0)})
= i (k\calD_{\infty_{-}} + m(k)\calD_{\infty_{+}}
+\calD_{\hat{\ul \mu} (x_0)})=0$,
the Riemann-Roch
theorem (\ref{a42}) implies
$r(-k\calD_{\infty_{-}}
- m(k)\calD_{\infty_{+}}
- \calD_{\hat{\ul \mu}(x_0)})= k+m(k)+1$
proving (\ref{b3}).
\end{proof}

Replacing $\phi$ by $\phi^{-1}$ one can discuss
$\calL(-k \calD_{\infty_{+}} - m(k)\calD_{\infty{-}}
- \calD_{\hat{\ul \nu}(x_0)} )$,
$k\in \bbN_0$ in an analogous fashion.

\vspace{6mm}
{\it Acknowledgments.} We thank Gerald Teschl
for discussions.


\begin{thebibliography}{10}
\label{ref}
\bibitem{alb79} S.~I.~Al'ber,  {\em Investigation of equations of
Korteweg-de Vries type by the method of recurrence relations},
J.~London
Math. Soc. (2) \textbf{19}, 467--480 (1979). (Russian.)


\bibitem{alb81} S. I. Al'ber, {\em On stationary problems for
equations of
Korteweg-de Vries type},   Commun. Pure Appl. Math. \textbf{34},
259--272 (1981).

\bibitem{alb87} S. I. Al'ber and M. S. Al'ber,
 {\em Hamiltonian formalism for nonlinear Schr\"{o}dinger
equations and sine-Gordon equations},    J.~London
Math. Soc. (2) \textbf{36},
176--192 (1987).

\bibitem{fva} F. V. Atkinson, {\em Discrete and Continuous
Boundary Problems}, Academic Press, New York, 1964.

\bibitem{bbei94}
E.~D.~Belokolos, A.~I.~Bobenko, V.~Z.~Enol'skii,
A.~R.~Its, and
V.~B.~Matveev, {\it Algebro-geometric Approach to
Nonlinear Integrable
Equations\/}, Springer, Berlin, 1994.

\bibitem{bght} W. Bulla, F. Gesztesy, H. Holden, and G. Teschl,
{\em
Algebro-geometric quasi-periodic finite-gap solutions of the
Toda and
Kac-van Moerbeke hierarchies}, Memoirs Amer. Math. Soc., to
appear.

\bibitem{bc1}  J. L. Burchnall and T. W. Chaundy, {\em Commutative
ordinary
differential operators},   Proc. London Math. Soc. (2)
\textbf{ 21}, 420--440 (1923).

\bibitem{bc2} J. L. Burchnall and    T. W. Chaundy, {\em
Commutative ordinary
differential operators},    Proc. Roy. Soc. London \textbf{A118},
557--583 (1928).

\bibitem{bc3} J. L. Burchnall and   T. W. Chaundy, {\em Commutative
ordinary
differential operators II. The identity
$P^n=Q^m$},  Proc. Roy. Soc. London
\textbf{A134}, 471--485 (1932).

\bibitem{ceek95}
P. L. Christiansen, J. C. Eilbeck, V. Z. Enolskii, and
N. A. Kostov, {\it Quasi-periodic solutions of the coupled
nonlinear Schr\"{o}dinger equations\/}, Proc. Roy. Soc. London
\textbf{A 451}, 685--700 (1995).

\bibitem{dj}
C. De Concini and R. A. Johnson, {\it The algebraic-geometric
AKNS potentials\/}, Ergod. Th. \& Dynam. Sys.
\textbf{7}, 1--24 (1987).

\bibitem{dickeyb}  L.A. Dickey, {\em Soliton Equations and
Hamiltonian
Systems},  World Scientific, Singapore, 1991.

\bibitem{dgu} R. Dickson, F. Gesztesy, and K. Unterkofler, {\em
A new approach to the Boussinesq hierarchy}, preprint, 1996.

\bibitem{dub77}
B. A. Dubrovin, {\it Completely integrable Hamiltonian systems
associated
 with matrix operators and Abelian varieties\/},
Funct. Anal. Appl.
\textbf{11}, 265--277 (1977).

\bibitem{dub83}
B. A. Dubrovin, {\it  Matrix finite-zone operators\/},
Revs. Sci. Tech.
\textbf{23}, 20--50 (1983).

\bi{dkn90}
B.~A.~Dubrovin, I.~M.~Krichever, and S.~P.~Novikov,
{\it Integrable
Systems. I\/}, in {\it Dynamical Systems IV\/},
V.~I.~Arnol'd and
S.~P.~Novikov (eds.), Springer, Berlin, 1990, pp. 173--280.

\bi{ef}
N.~M.~Ercolani and H.~Flaschka, {\it The geometry of the Hill
equation and of the Neumann system\/}, Phil. Trans. Roy. Soc.
London \textbf{A 315}, 405--422 (1985).

\bi{fk}
H.~M.~Farkas and I.~Kra, {\it Riemann Surfaces\/},
2nd ed., Springer,
 New York,
1992.

\bi{fay}
J.~D.~Fay, {\it Theta Functions on Riemann Surfaces\/},
Lecture Notes
in Mathematics 
{\bf 352},
Springer, Berlin, 1973.

\bibitem{gd79} I. M. Gel'fand  and L. A. Dikii, {\em Integrable
nonlinear
equations and the Liouville theorem}, Funct. Anal. Appl.
\textbf{13}, 6--15
(1979).

\bi{ges92}
F.~Gesztesy, {\it Quasi-periodic, finite-gap solutions
of the modified
Korteweg-de Vries equation}, in {\it Ideas and
Methods in Mathematical
Analysis, Stochastics, and Applications,
Volume 1\/}, S.~Albeverio,
J.~E.~Fenstad, H.~Holden, and T.~Lindstr\o m (eds.),
Cambridge University
Press, Cambridge, 1992, pp.~428--471.

\bi{gs}
F.~Gesztesy and R.~Svirsky, {\it (m)KdV solitons on
the background of
quasi-periodic finite-gap solutions\/}, Memoirs
Amer.\ Math.\ Soc.\,
\textbf{118}, No. 563
(1995).

\bibitem{gw93} F.~Gesztesy,   R.~Weikard.  {\em  Spectral
deformations and
soliton equations}, in {\it Differential Equations with
Applications to
Mathematical Physics}, W.~F.~Ames, E.~M.~Harrell~II, and
J.~V.~Herod (eds.), Academic Press, Boston, 1993, pp.~101--139.

\bibitem{geweMN} F. Gesztesy and R. Weikard, {\em Lam{\'{e}}
potentials and
the stationary (m)KdV hierarchy},  Math. Nachr.
\textbf{176}, 73--91 (1995).

\bibitem{geweMZ}    F. Gesztesy and R. Weikard, {\em
Treibich-Verdier
potentials and the stationary (m)KdV hierarchy},  Math. Z.
\textbf{219}, 451--476 (1995).


\bibitem{geweAM}  F.~Gesztesy and R.~Weikard, {\em  Picard
potentials and
Hill's equation on a torus}, Acta  Math. {\bf 176}, 73--107
(1996).

\bi{grt} F.~Gesztesy, R.~Ratnaseelan, and  G.~Teschl, {\em The
KdV hierarchy
and associated trace formulas}, in proceedings of the
{\it International Conference on Applications of Operator Theory},
I.~ Gohberg, P.~ Lancaster, P.~ N.~ Shivakumar (eds.),  Operator
Theory: Advances and Applications, Vol. 87, Birkh\"{a}user, 1996,
pp.~125--163.

\bibitem{grgu}  B.~Grebert and J.~C.~Guillot, {\em
Gaps of one dimensional periodic AKNS System},
Forum Mathematicum, to appear.

\bi{gh}
P.~Griffiths and J.~Harris, {\it Principles of Algebraic Geometry\/},
Wiley, New York, 1978.

\bibitem{its1}    A. R. Its, {\em  Inversion of hyperelliptic
integrals
 and integration of nonlinear differential equations},
Vestnik Leningrad Univ. Math.
\textbf{9}, 121--129 (1981).

\bibitem{its2}    A. R. Its, {\em  On Connections
 between solitons and finite-gap solutions of the
nonlinear Schr\"{o}dinger equation},
Sel. Math. Sov.
\textbf{5}, 29--43 (1986).

\bibitem{itma}  A. R. Its and V. B. Matveev,
{\em Schr\"odinger
operators with finite-gap spectrum and N-soliton solutions of the
Korteweg-de Vries equation}, Theoret. Math. Phys. \textbf{23},
343--355
(1975).

\bibitem{jac}
C.~G.~T.~Jacobi, {\em \"Uber eine neue  Methode zur Integration der
hyperelliptischen Differentialgleichungen  und \" uber die
rationale Form
ihrer vollst\" andigen algebraischen Integralgleichungen}, J.~Reine
Angew.  Math.  \textbf{32}, 220--226  (1846).

\bi{kra}
A.~Krazer, {\it Lehrbuch der Thetafunktionen\/},
Chelsea, New
York, 1970.

\bi{kri1}
I.~M.~Krichever, {\it Integration of nonlinear
equations by the methods of algebraic geometry\/},
 Funct.\ Anal.\ Appl.\ {\bf
11}, 12--26 (1977).

\bi{kri2}
I.~M.~Krichever, {\it Nonlinear equations and elliptic curves\/},
 Revs. Sci. Tech. {\bf
23}, 51--90 (1983).

\bi{ma}
Y.--C. Ma and M. J. Ablowitz,
{\it The periodic cubic Schr\"odinger equation\/},
  Stud.\ Appl.\ Math.\  {\bf
65}, 113--158 (1981).

\bibitem{mk85}
H.~P.~McKean, {\em Variation on a theme of
Jacobi}, Commun.  Pure Appl.  Math.  \textbf{38}, 669--678 (1985).

\bi{mer}
J. Mertsching,
{\it Quasi periodic solutions of the
nonlinear Schr\"odinger equation\/},
  Fortschr.\ Phys.\  {\bf
85}, 519--536 (1987).

\bibitem{mum1}  D.~Mumford, {\em Tata Lectures on Theta I},
Birkh\"auser, Boston, 1983.

\bibitem{mum2}  D.~Mumford, {\em Tata Lectures on Theta II},
Birkh\"auser, Boston, 1984.

\bibitem{new}  A. C. Newell, {\em Solitons in Mathematics
and Physics },
SIAM, Philadelphia, 1985.

\bi{nmpz}
S.~Novikov, S.~V.~Manakov, L.~P.~Pitaevskii, and
V.~E.~Zakharov, {\it
Theory of Solitons\/}, Consultants Bureau,
New York, 1984.

\bibitem{pre85} E. Previato, {\em  Hyperelliptic quasi-periodic
 and soliton solutions of the nonlinear Schr\"odinger equation\/},
Duke  Math. J. \textbf{52}, 329--377 (1985).

\bibitem{pre96} E. Previato, {\em Seventy years of spectral curves:
1923--1993\/}, in {\it Integrable Systems and Quantum Groups\/},
M.~Francaviglia and S.~Greco (eds.), Lecture Notes in Math.,
Vol.~1620, Springer, Berlin, 1996,
pp.~419--481.

\bi{sch}
R.~Schimming, {\it An explicit expression for the Korteweg--de Vries
hierarchy\/}, Acta Appl. Math. \textbf{39}, 489--505 (1995).

\bibitem{smi95} A. O. Smirnov, {\em Elliptic solutions of the
nonlinear Schr\"odinger equation and the modified
 Kortweg-de Vries  equation\/},
Russian Acad. Sci. Sb. Math.   \textbf{82}, 461--470 (1995).

\bibitem{wi85} G. Wilson, {\em Algebraic curves and soliton
equations},  in
{\it Geometry Today\/},  E.~Arbarello, C.~Procesi, and E.~Strickland
(eds.),
Birkh\"auser, Boston, 1985, pp.~303--329.


\end{thebibliography}
\end{document}